\documentclass[]{raa}            
\usepackage{natbib}
\usepackage{times}

\usepackage{amssymb}
\usepackage{amsfonts}
\usepackage{mathrsfs}
\usepackage[dvips]{graphicx}

\providecommand{\bm}[1]{\mbox{\boldmath$#1$\unboldmath}}

\begin{document}

   \title{Surface photometry and radial color gradients of nearby luminous early-type galaxies in SDSS Stripe 82
$^*$
\footnotetext{\small $*$ Supported by the National Fund for Fostering Talents of Basic Sciences of China.}
}

 \volnopage{ {\bf 2010} Vol.\ {\bf XX} No. {\bf X}, XX--XX}
   \setcounter{page}{1}

   \author{Fang-Zhou Jiang
      \inst{1,2}
   \and Song Huang
      \inst{2}
   \and Qiu-Sheng Gu
      \inst{2}
   }

   \institute{Department for Intensive Instruction, Kuang Yaming Honors School, Nanjing University,
             Nanjing 210093, China; {\it fangzhou.jiang@yale.edu}\\
        \and
             Department of Astronomy, Nanjing University, Nanjing 210093, China
\vs \no \\
   {\small Received 2010 July 8; accepted 2010 September 13 }
}

\abstract{ We make use of the images from the Sloan Digital Sky Survey Stripe 82  (Stripe 82) to present an analysis of $r$ band surface brightness profiles and radial  color gradients ($g-r$, $u-r$) in our sample of 111 nearby early-type galaxies (ETGs). Thanks to the Stripe 82 images, each of which is co-added from about 50 single frames, we are able to pay special attentions to the low-surface-brightness areas (LSB areas) of the galaxies. The LSB areas make a difference to the S\'{e}rsic fittings and concentration indices, making both the indices less than the typical values for ETGs. In the S\'{e}rsic fits to all the surface brightness profiles, we found some S\'{e}rsic indices range from 1.5 to 2.5, much smaller than that of typical de Vaucouleur profiles and relatively close to that of exponential disks, and some others much larger than 4 but still with accurate fitting. Two galaxies cannot be fitted with single S\'{e}rsic profile, but once we try double S\'{e}rsic profiles, the fittings are improved: one with a profile relatively close to de Vaucouleur law in the inner area and a profile relatively close to exponential law in the LSB area, the other with a nice fitting in the inner area but still a failed fitting in the outer area. There are about 60\% negative color gradients (red-core) within 1.5$R_{\textrm{\footnotesize{e}}}$, much more than the approximately 10\% positive ones (blue-core) within the same radius. However, taking into account of the LSB areas, we find that the color gradients are not necessarily monotonic: about one third of the red-core (or blue-core) galaxies have positive (or negative) color gradients in the outer areas. So LSB areas not only make ETGs' S\'{e}rsic profiles deviate from de Vaucouleur ones and shift to the disk end, but also reveal that quite a number of ETGs have opposite color gradients in inner and outer areas. These outcomes remind us the necessity of double-S\'{e}rsic fitting. These LSB phenomena may be interpreted by mergers and thus different metallicity in the outer areas. Isophotal parameters are also discussed briefly in this paper: more disky nearby ETGs are spotted than boxy ones.
}

   \authorrunning{F. Z. Jiang,  S. Huang \& Q. S. Gu }            
   \titlerunning{Surface Photometry and Color Gradients of Nearby ETGs in SDSS Stripe 82}  
   \maketitle


%
%
\section{Introduction}           
\label{sect:intro}

Early-type galaxies (ETGs), including elliptical galaxies (Es) and bulge-dominated S0 galaxies (S0s), are generally known to be dynamically simple stellar systems of homogeneous stellar population, devoid of dust, cool gas and young blue stars. However, they are far from relaxation and their morphological variety (e.g., Es' triaxial/oblate shapes and S0s' disks) implies that they originated by different means. The spatial distribution of stellar population properties in ETGs are the chemodynamical fossil imprints
of galaxy formation and evolution mechanisms. The study of surface brightness profiles and radial color profiles of ETGs sheds light on the star formation history and buildup of the stellar population, and thus provides important clues to understand galaxy formation and evolution. 

As a generalization to de Vaucouleur's $R^{\frac{1}{4}}$ law, S\'{e}rsic's (1963, 1968) $R^{\frac{1}{n}}$ model is widely used to describe the stellar distributions in galaxies. The stellar distributions of ETGs are equivalent to the surface brightness distributions in terms of optical observations, including Sloan Digital Sky Survey (SDSS). Even early-type stellar systems are somewhat mixtures of bulge and disk components, which combines to result in the intermediary form between $R^{\frac{1}{4}}$ bulge and $R^{\frac{1}{1}}$ disk, i.e., the $R^{\frac{1}{n}}$ profile. Now except for decomposing an image into its separate components, galaxies are modeled with a single S\'{e}rsic profile (e.g., Blanton et al. 2003). 

The colors of ETGs get bluer from the center outwards (Boroson et al. 1983; Kormendy \& Djorgovski 1989; Franx \& Illingworth 1990; Peletier et al. 1990a, 1990b; Michard 1999; Idiart et al. 2002; de Propris et al. 2005; Wu et al. 2005; La Barbera \& De Carvalho 2009). The negative color gradients are believed to be metallicity-dominated (Ferreras et al. 2009, Spolaor et al. 2010) in spite of the famous age-metallicity degeneracy (Worthey 1994), although whether or not they evolve with cosmic time is still under debate (e.g., Hinkley \& Im 2001). Classic collapse model (Eggen et al. 1962; Larson 1974a, 1974b, 1975; Carlberg 1984) suggests that ETGs form in high redshift without subsequent secondary star formation, while others (Hinkley \& Im 2001) indicate that secondary bursts/accretions may occur at moderate redshifts. Recent studies discovered a significant fraction of positive color gradients (Michard 1999, Menanteau et al. 2001, Ferreras et al. 2005, Elmegreen et al. 2005, Suh et al. 2010), which have something to do with age gradients in addition to metallicity gradient (Silva \& Elston 1994, Michard 2005). Age gradients could result from galaxy mergers (e.g., Toomre \& Toomre 1972) where star formation events continue or occur episodically. Actually, a remarkable fraction of ETGs at high redshifts exhibit positive color gradients as results of mergers or inflows (Menanteau et al. 2001, Marcum et al. 2004, Ferreras et al. 2005, Elmegreen et al. 2005). This phenomena suggest that secondary star formation events take place at the centers of these ETGs. In consistent with these results, some post-starburst galaxies such as E+A galaxies have positive color gradients, which then evolve into negative ones as the population ages (e.g., Yang et al. 2008).

This paper presents results from surface photometry of 111 nearby luminous ETGs from SDSS Stripe 82 (Frieman et al. 2008, Sako et al. 2008) in three bands, $u$, $g$ and $r$. We use S\'{e}rsic profiles to fit their $r$ band surface brightness profiles, and we measure their color profiles of $g-r$ and $u-r$. Benefitted from the co-added images of SDSS Stripe 82, we pay attention to the low-surface-brightness areas (LSB areas) of the ETGs and influences from the LSB areas to the S\'{e}rsic fits and the radial color gradients. Section 2 describes the sample and the data reduction. Section 3 presents the results of the S\'{e}rsic fits, the analyses of color gradients and isophotal parameters. Section 4 is a brief discussion for the previous section. Finally, we summarize the conclusions in Section 5. Except where stated otherwise, we assume a $\Lambda$CDM cosmology with $\Omega _{\textrm{\footnotesize{m}}}=0.3$ and $H_{\textrm{\scriptsize{0}}}=70\textrm{kms}^{-1}\textrm{Mpc}^{-1}$. 

\section{Sample and data reduction}
\label{sect:sample}

\subsection{ETG Selection in SDSS Stripe 82}
SDSS Stripe 82 is part of the SDSS-II supernova survey (Frieman et al. 2008, Sako et al. 2008). A 2.5 degree wide region along the celestial equator from $-59<$RA$<59$ is imaged repeatedly in 303 runs (plus 2 coadd runs) for three months (September, October and November) in each of three years (2005-2007). Each image is co-added from ~50 ordinary SDSS images of the same object, each of which has the exposure time of 52 seconds. So the co-added Stripe 82 images are about 2 magnitudes deeper than single frames. With the extra depth, some features invisible in single images are revealed in the LSB outer areas. Fig. 1 shows two examples of the LSB structures and Table 1 gives the information of the two ETGs (in our sample). 

\begin{figure}[h!!!]
\centering
\includegraphics[width=9.0cm, angle=90]{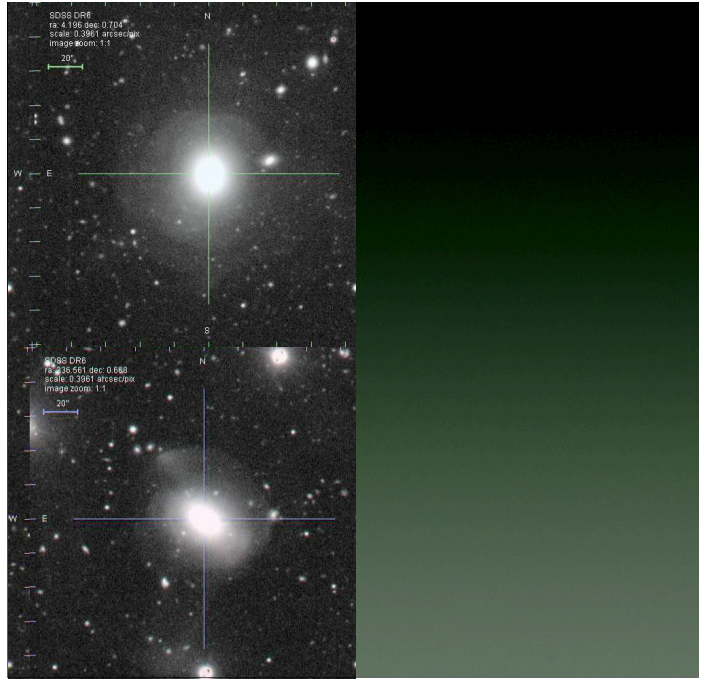}
\begin{minipage}[]{85mm}
\caption{ Left: Stripe 82 images showing shell arches or interactions in LSB areas. \quad Right: ordinary SDSS images for the same galaxies on their left.} 
\end{minipage}
\label{fig1}
\end{figure}

\begin{table}[h!!!]
\small
\centering
\begin{minipage}[]{120mm}
\caption[]{ Information of the Two ETGs whose Stripe 82 Images Show LSB Features}\label{tab1}\end{minipage}
\tabcolsep 6mm
 \begin{tabular}{clclc}
  \hline\noalign{\smallskip}
SDSS Name &  RA      & DEC & $r$ Magnitudes  & $r$ FracDev              \\
  \hline\noalign{\smallskip}
J001647.00+004215.3  & 4.19587 &  0.70426    & 14.04 & 0.95 \\ 
J222614.62+004004.1  & 336.56095 &  0.66783    & 14.74 & 1 \\ 
  \noalign{\smallskip}\hline
\end{tabular}
\end{table}

These are two examples of our sample of 111 ETGs, the vast majority of which are chosen from a larger sample of more than 400 galaxies according to the following criteria: Petrosian $r<15$, $z<0.02$ and FracDev$_r>0.9$, which mean that they are all strictly nearby luminous early-type stellar systems. FracDev indicates how close a galaxy is in accordance with de Vaucouleur profile: 1 means typical de Vaucouleur profile. The larger sample of more than 400 luminous galaxies are chosen merely according to $r<15$, so it includes irregular ones, spiral ones and early-type ones. Then eyeball review adds some obvious ETGs into the sample, even if some of them has FracDev$_r<0.9$.

\subsection{1D Surface Photometry with S\'{e}rsic Fits and Measurement of Color Gradients}
All the images that we used are corrected by the SDSS photometric pipeline (flat-fielded, sky-subtracted, etc..). SDSS pipeline may sometimes overestimate the sky background, but our scientific purpose is not about the absolute values of the surface brightness. The $2046\times1489$-pixel frames are sampled with $0''.396\times 0''.396$ pixels. The PSF in $r$ band is also $0''.396$. The zero points of photometry are 24.80, 25.11 and 24.63 in $r$, $g$ and $u$ band respectively. 

We have 111 images in each band and thus 333 frames in all the three wavebands. For each of the 333 images, now that the sky background has been subtracted, we use SExtractor to generate segment image to mask all the sources other than the target. In order to have a neat and complete mask, we adopt the 'automatic+manual' scheme. First, we expand the segments appropriately, leaving the target and the sources close to the target not masked. Second, we mask any contaminations on target as well as any other sources close to our target manually in the interactive mode of ELLIPSE. ELLIPSE is the 1D surface photometry task in the STSDAS ISOPHOTE package in Image Reduction Facility (IRAF).\footnote{http://iraf.noao.edu/} 

Secondly, we measure the surface brightness along elliptical annuli in $r$ band using ELLIPSE. ELLIPSE is run twice for $r$ band. We adopt the coordinates of the centers of the galaxies provided by SDSS, since we have tested their accuracy one by one by an extra run of ELLIPSE and find that the average values are no more than one pixel different from the SDSS ones. So in the first run, the centers of all the isophotes are fixed, while the position angle (PA) and the ellipticity ($e$) are set free as functions of radius, as well as the surface brightness ($\mu$). The steps of the semi-major axis (SMA) are nonlinear within the upper limit of 800 pixels (ELLIPSE parameter maxsma=800). Neighboring isophotes may contact or cross each other in the first run, when the 'maximum sma length for iterative mode' (ELLIPSE parameter maxrit) is the same with 'maxsma'. So in the second run, 'maxrit' is reduced to such an extent that no neighboring isophotes contact or cross each other, which can be tell by the PA profile and the $e$ profile -- they should vary continuously without any jump or break. We then convert the photometric table to a spreadsheet and import its useful columns (SMA, INTENS, ELLIP, MAG, MAG\_LERR, MAG\_UERR, A4) into MATLAB to calculate isophotal parameters $e$, $a_4/a$ and to plot the surface brightness profile. \footnote{The meaning of $a_4/a$ and how to calculate $e$ and $a_4/a$ are discussed in Section 3.} We apply single S\'{e}rsic fit to the profile to give the least-squares estimate of the three fitting parameters: S\'{e}rsic index $n$, effective radius $R_{\textrm{\footnotesize{e}}}$ and $\mu_{\textrm{\footnotesize{e}}}$, the surface brightness within $R_{\textrm{\footnotesize{e}}}$. We use this $R_{\textrm{\footnotesize{e}}}$ as $R_{\textrm{\scriptsize{50}}}$ and adopt $R_{\textrm{\scriptsize{90}}}$ from SDSS) to calculate concentration index in $r$ band, $C_r=R_{\textrm{\scriptsize{90}}}/R_{\textrm{\scriptsize{50}}}$. Please note that before fitting, the innermost 2PSF region of the profile is truncated as well as the outermost end of the profile where the brightness approaches the sky level. So even if the innermost pixels are saturated, it does not matter.

Thirdly, we apply the same set of isophotes from the photometry in $r$ band to the photometry in $u$ and $g$ bands so that the surface brightness profiles in all the three bands have exactly the same steps of radii. Hence, simple subtractions between surface brightness profiles give radial color profiles. We divide each color profile into two regions: the inner region 0.5$R_{\textrm{\footnotesize{e}}}$--1.5$R_{\textrm{\footnotesize{e}}}$ and the outer LSB region 1.5$R_{\textrm{\footnotesize{e}}}$--4$R_{\textrm{\footnotesize{e}}}$. The former region is for the convenience of comparing our analyses on color gradients with others' works (e.g., Suh et al. 2010), for previous statistical works discuss color profiles in the same region more or less: within 0.5$R_{\textrm{\footnotesize{e}}}$ is approximately the region affected by seeing, while beyond 1.5$R_{\textrm{\footnotesize{e}}}$, single SDSS images can hardly give enough signal-to-noise ratio. The 1.5$R_{\textrm{\footnotesize{e}}}$--4$R_{\textrm{\footnotesize{e}}}$ region is for our emphases on the LSB areas, beyond which noise dominates.

\section{Results}
\label{sect:results}

\subsection{S\'{e}rsic Fits and Concentration Indices}
S\'{e}rsic's (1963, 1968) $R^{\frac{1}{n}}$ model is commonly expressed as an intensity profile (e.g., Sparke \& Gallagher 2007).
\begin{equation}
I(R)=I_{\textrm{\scriptsize e}}e^{-b_n[(\frac{R}{R_{\textrm{\tiny e}}})^{\frac{1}{n}}-1]},
\label{eq:sersic1}
\end{equation}
where $I_{\textrm{\footnotesize{e}}}$ is the intensity at effective radius $R_{\textrm{\footnotesize{e}}}$, and $b_n$ is selected to make sure
\begin{equation}
\int_{0}^{\infty}I(R)2\pi RdR=2\int_{0}^{R_{\textrm{\tiny e}}}I(R)2\pi RdR,
\label{eq:sersic2}
\end{equation}
so that the effective radius $R_{\textrm{\footnotesize{e}}}$ encloses half of the total light from the model (Ciotti 1991, Caon et al. 1993). However, the photometric results are in the units of magnitude, which by definition should be converted from intensity with the formula
\begin{equation}
\mu(R)=-2.5\log I(R).
\label{eq:sersic3}
\end{equation}
So the actual format of the S\'{e}rsic's empirical formula that are used in fitting is 
\begin{equation}
\mu(R)=\mu_{\textrm{\footnotesize{e}}}+\frac{2.5b_n}{\ln(10)}[(\frac{R}{R_{\textrm{\footnotesize e}}})^{\frac{1}{n}}-1],
\label{eq:sersic4}
\end{equation}
where
\begin{equation}
b_n=2n-\frac{1}{3}+\frac{4}{405n}+\frac{46}{25515n^2}+O(n^{-3}), n>0.36.
\label{eq:sersic5}
\end{equation}
The deduction and approximation of $b_n$ can be referred to Graham \& Driver (2005).

In Appendix Table A.1 and Table A.2, we list the photometric results. We do find quite a number of de Vaucouleur profiles (Fig. 2) with $n\approx 4$, but there are also several $1.5<n<2.5$ profiles which are relatively close to exponential disks (Fig. 3). Furthermore, a couple of profiles have S\'{e}rsic indices much larger than 4 but can still be fitted well with single S\'{e}rsic profile (Fig. 4), while only two  profiles cannot be fitted with single S\'{e}rsic at all (Fig. 5). However, once we try curve fitting with double S\'{e}rsic profiles, one of the two above-mentioned ones is fitted much better with an inner $n=4.68$ curve and an outer $n=1.45$ curve (Fig. 6 Left), which indicates that this ETG may be classified as elliptical galaxy with single SDSS image but actually turns out to be an S0 system once its LSB region is revealed. But the other one still cannot be fitted well even with double S\'{e}rsic profiles (Fig. 6 Right), no matter how we change the initial value of $n$, $\mu_{\textrm{\footnotesize{e}}}$ and $R_{\textrm{\footnotesize{e}}}$ or how we try different radii where the two curves meet. This failure indicates that the components of this ETG can hardly be distinguished with 1D S\'{e}rsic fit or fits. 

\begin{figure}
\includegraphics[width=75mm]{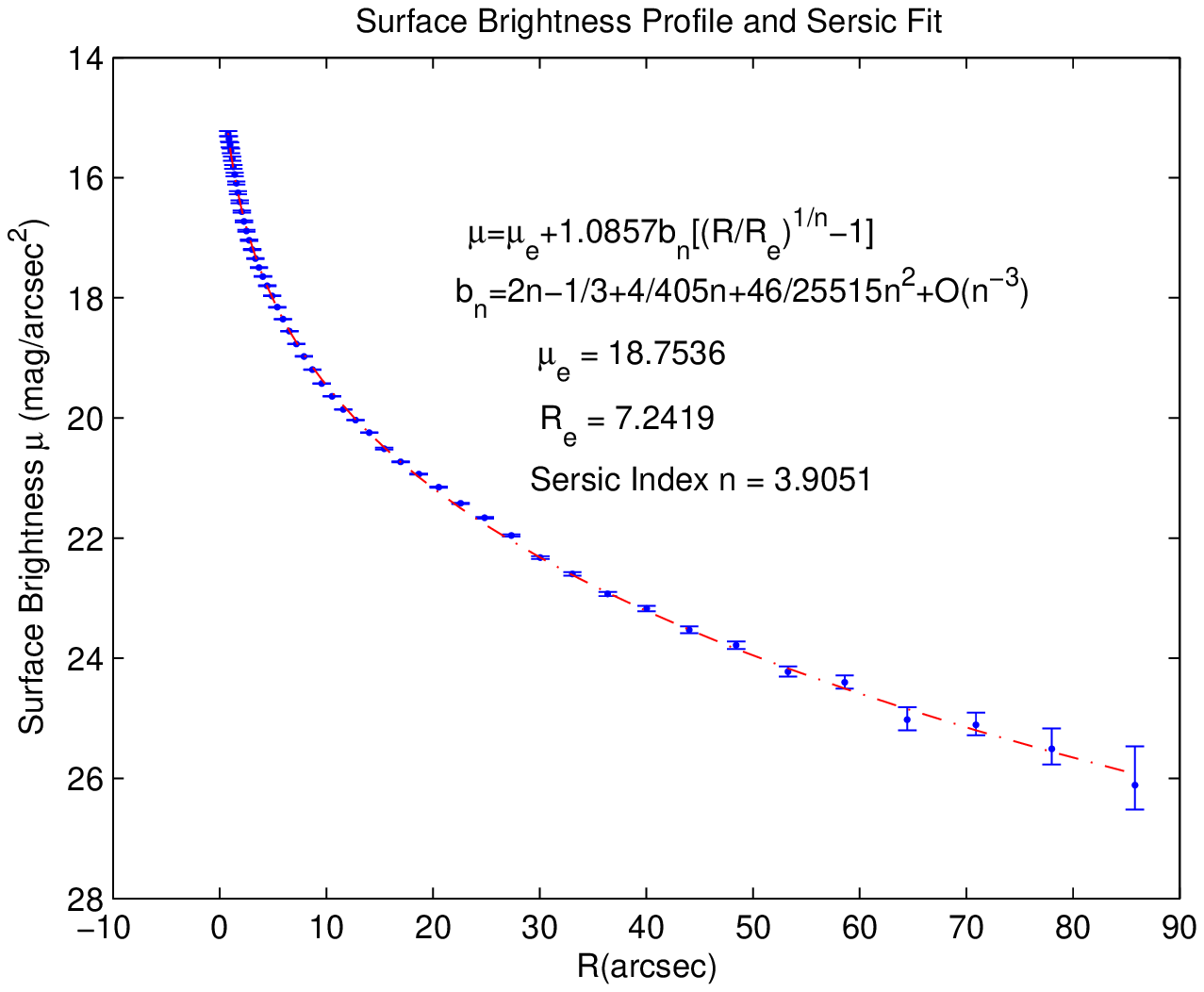}
\includegraphics[width=75mm]{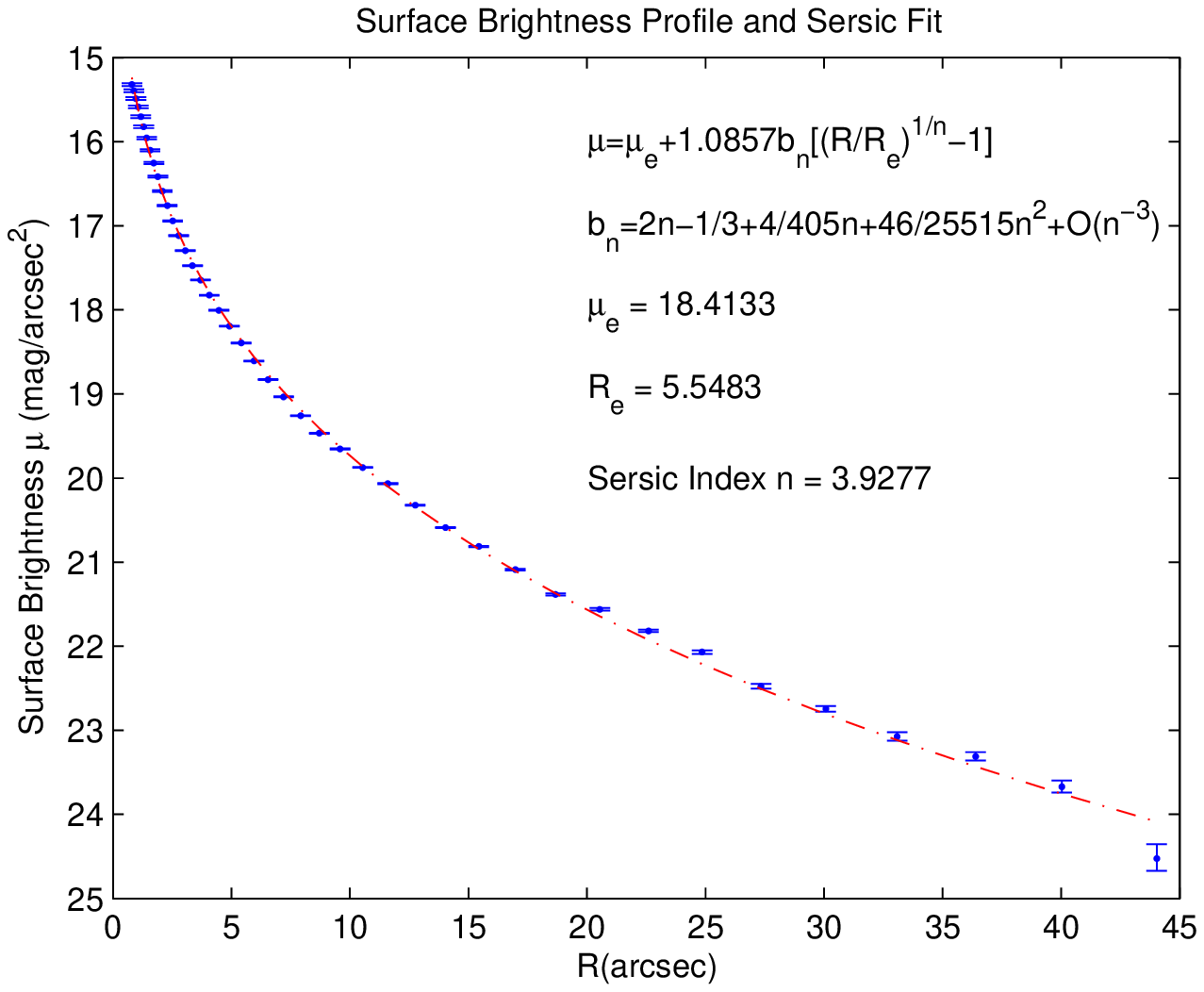}
\caption{Two examples of typical de Vaucouleur profiles, $n\approx 4$. The dashed line is the S\'{e}rsic's $R^{\frac{1}{n}}$ model. }
   \label{fig2}
\end{figure}

\begin{figure}
\includegraphics[width=75mm]{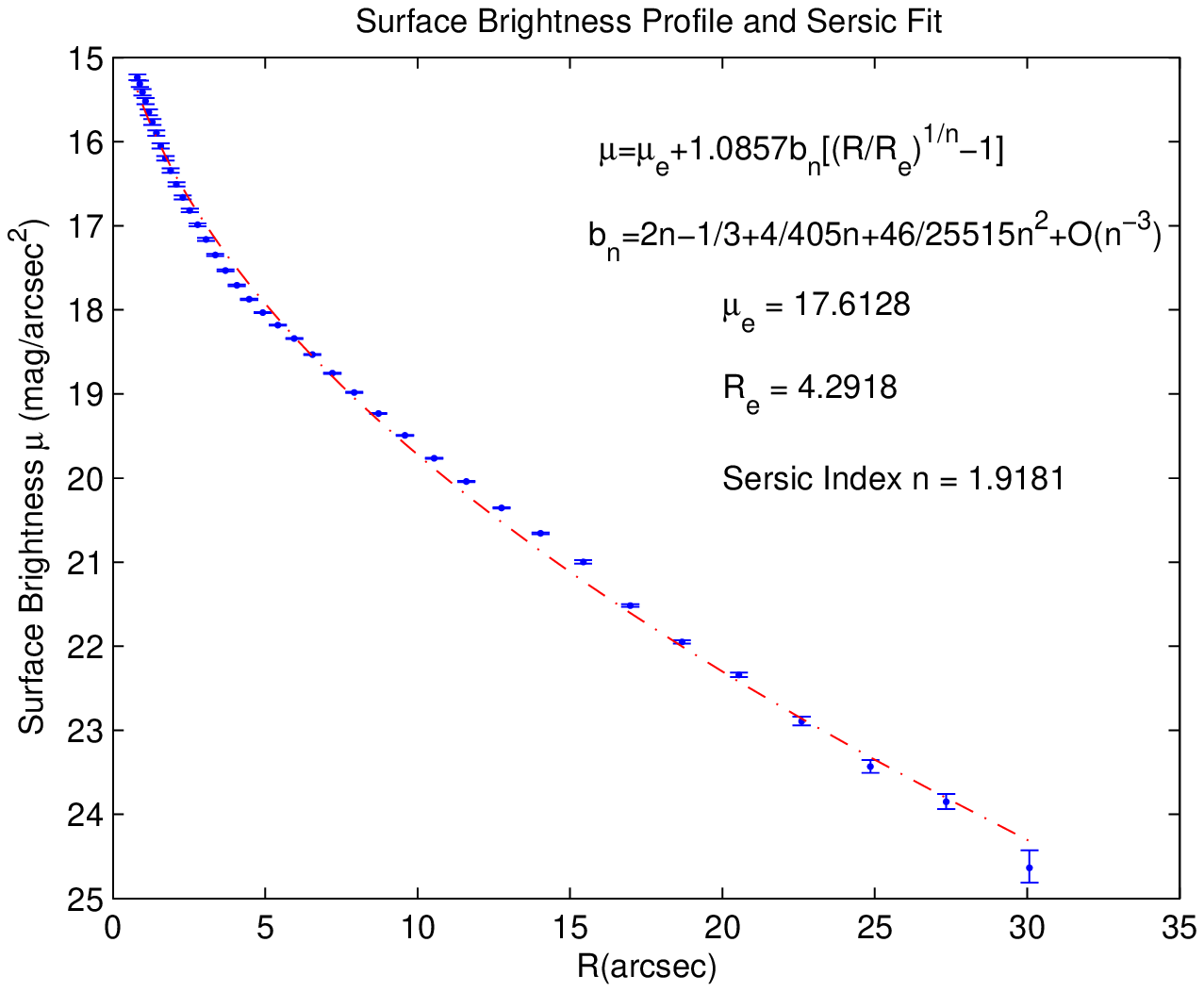}
\includegraphics[width=75mm]{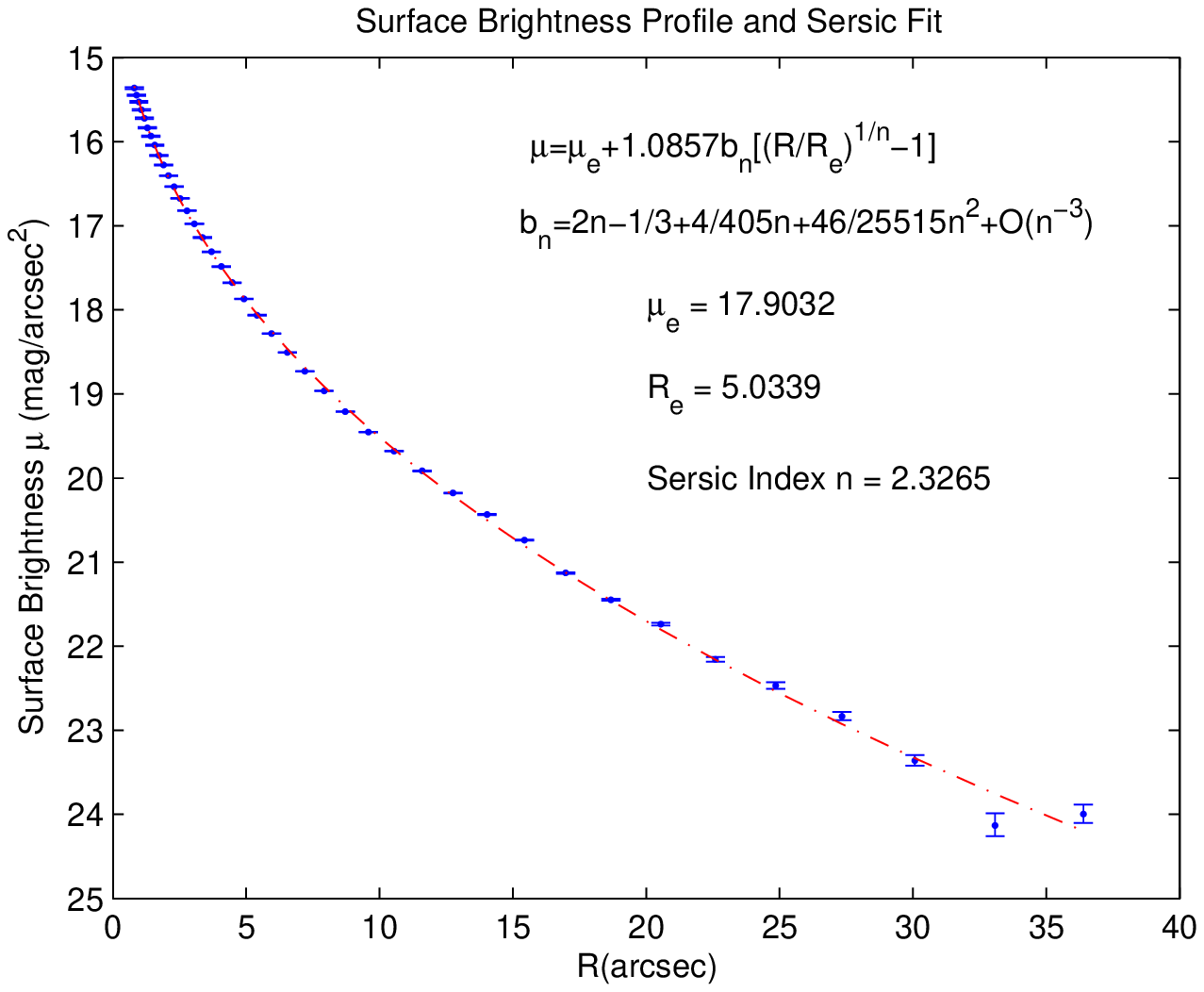}
\caption{Two examples of the $1.5<n<2.5$ profiles which are relatively close to exponential disks. The dashed line is the S\'{e}rsic's $R^{\frac{1}{n}}$ model. }
   \label{fig3}
\end{figure}

\begin{figure}
\includegraphics[width=75mm]{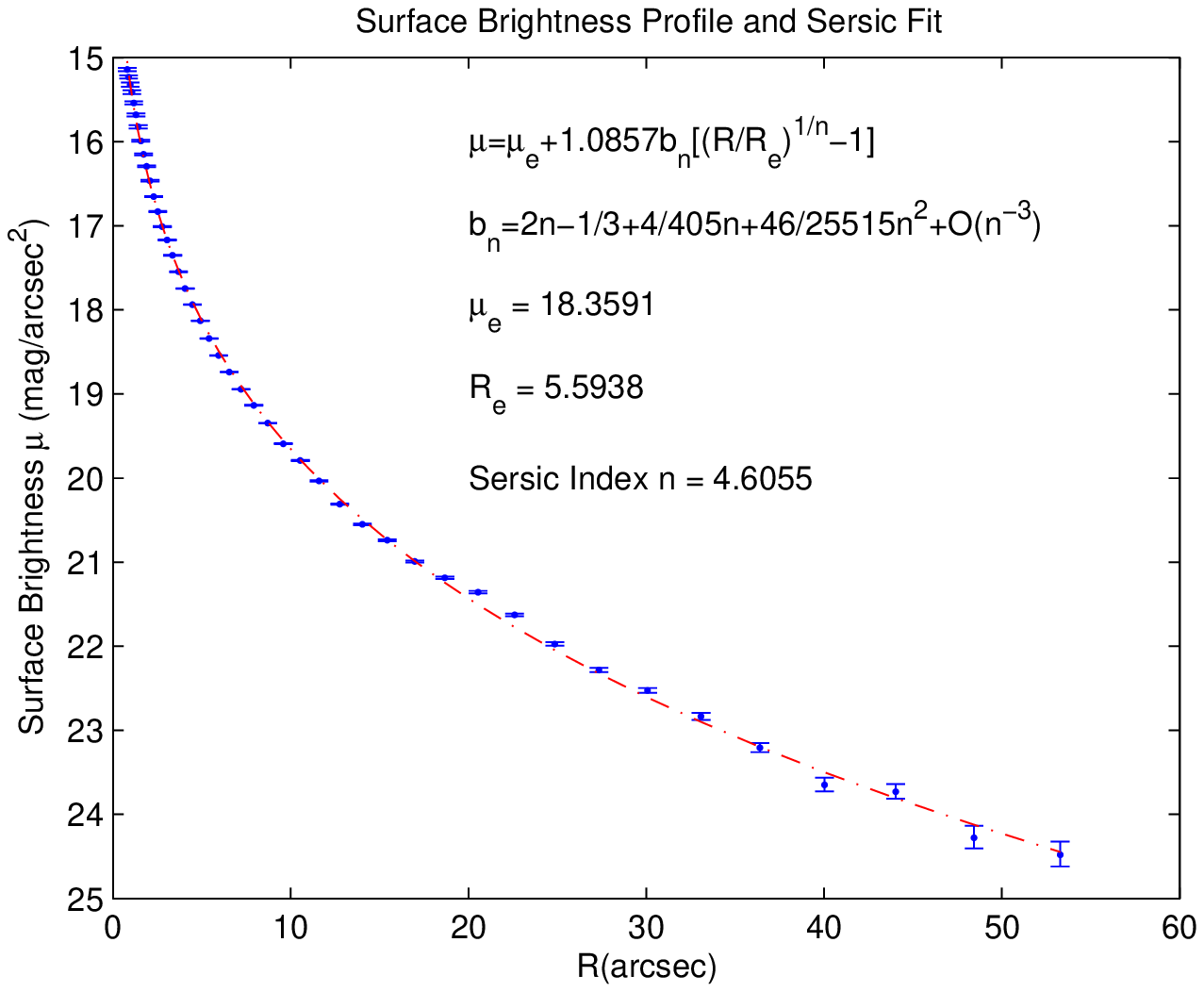}
\includegraphics[width=75mm]{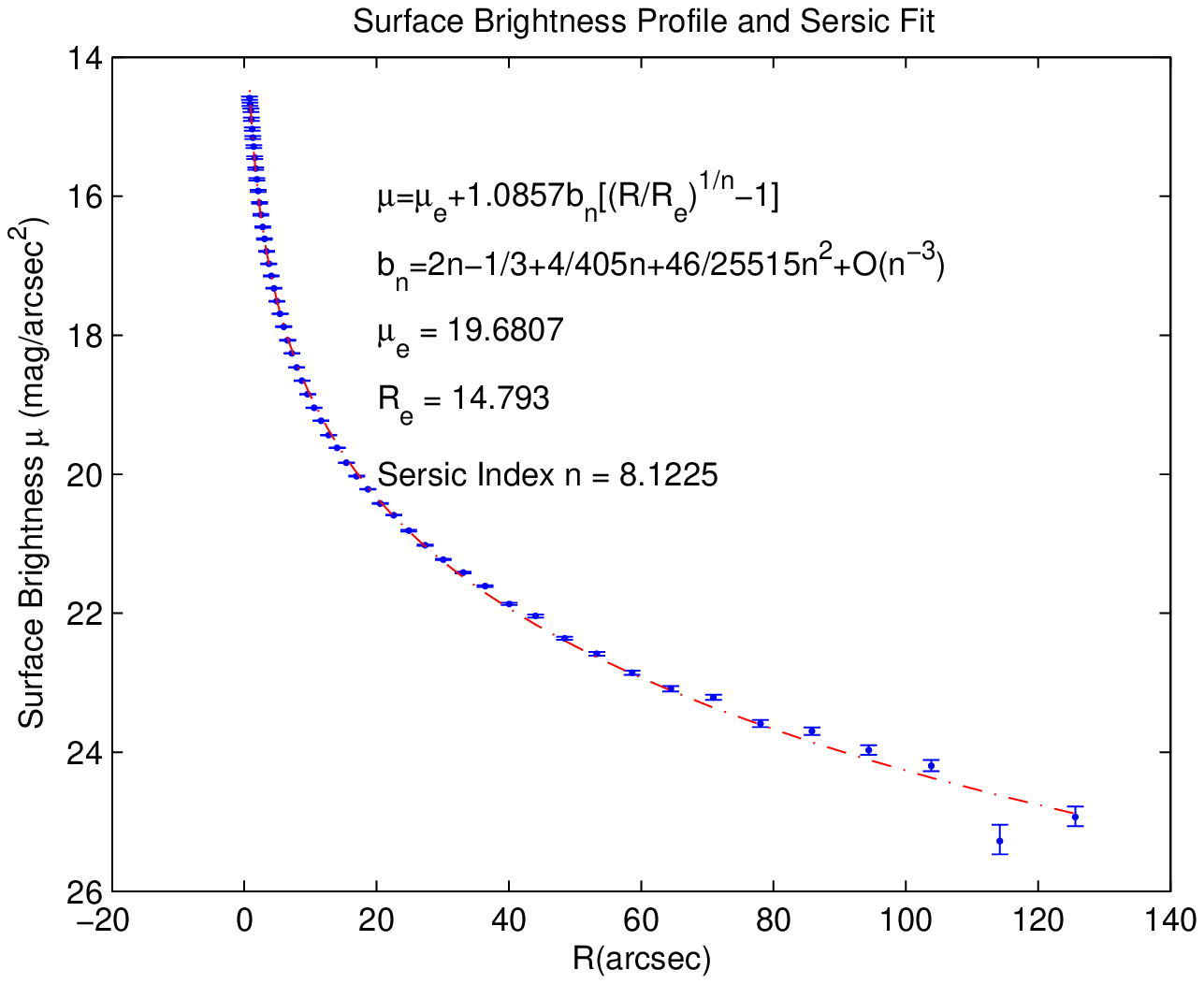}
\caption{Two examples of the well-fitted single S\'{e}rsic profiles even with $n$ much larger than 4. The dashed line is the S\'{e}rsic's $R^{\frac{1}{n}}$ model. }
   \label{fig4}
\end{figure}

\begin{figure}
\includegraphics[width=75mm]{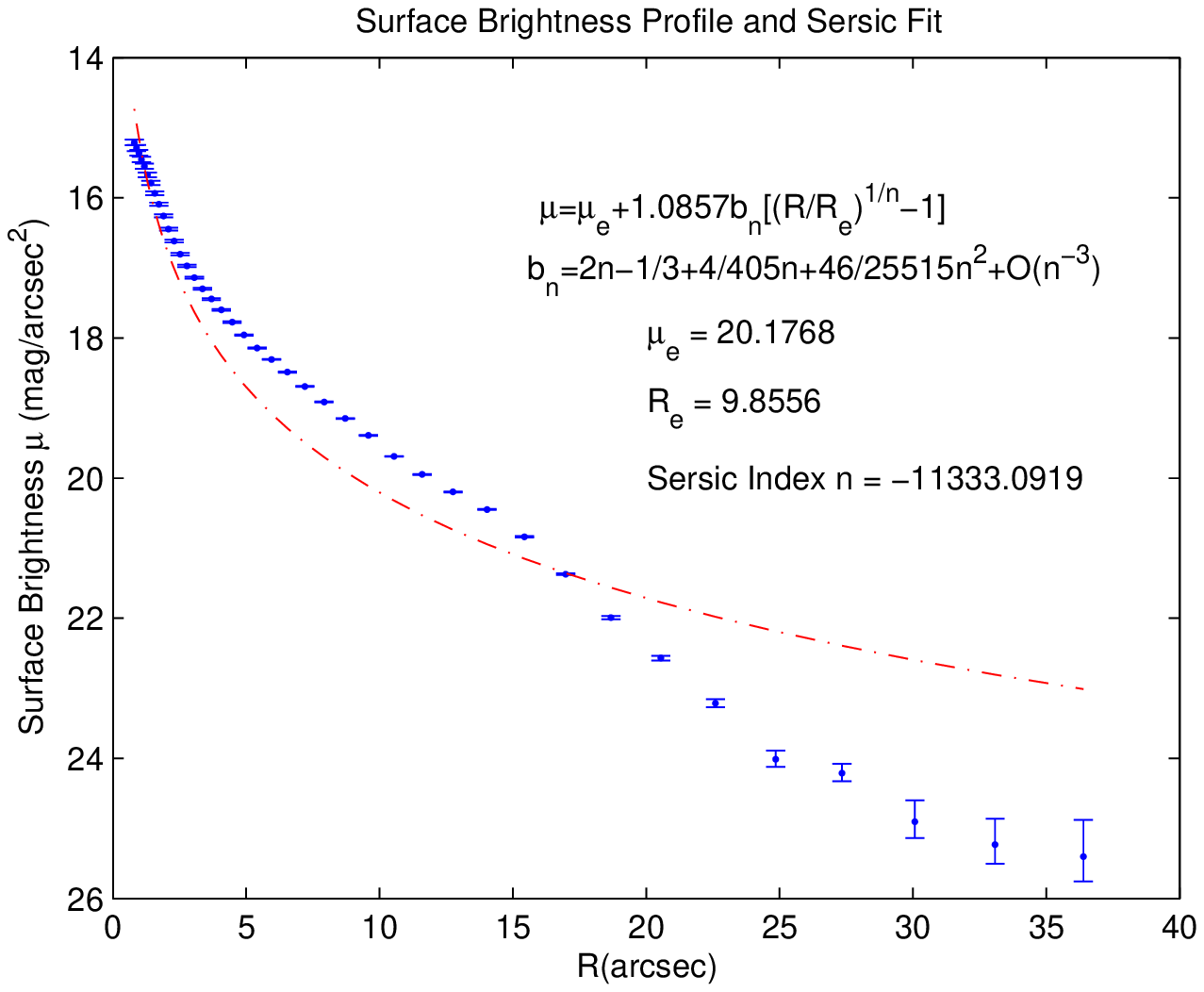}
\includegraphics[width=75mm]{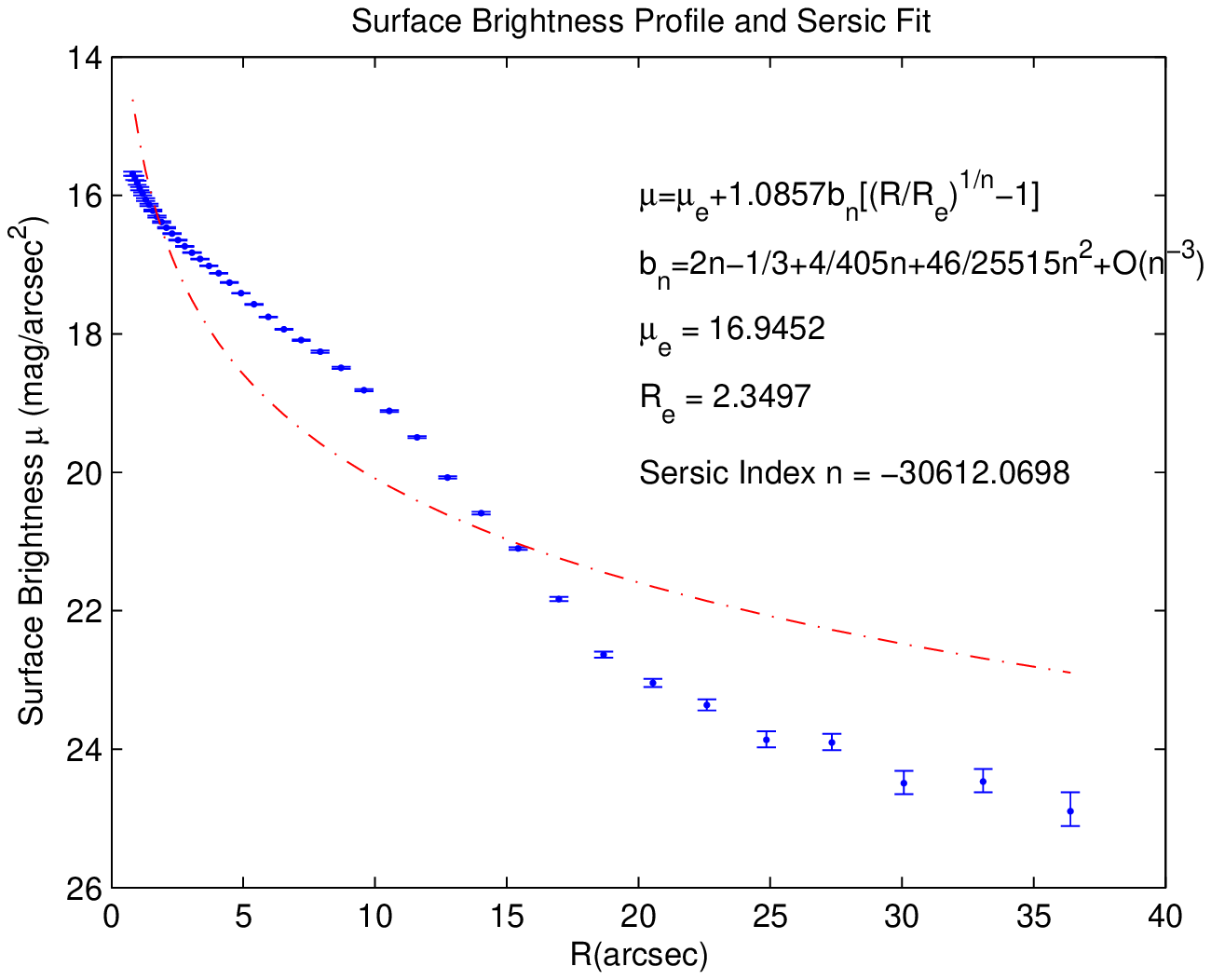}
\caption{Single S\'{e}rsic profile fails to fit any of the two. The dashed line is the S\'{e}rsic's $R^{\frac{1}{n}}$ model but the $n$ values here do not make any sense.}
   \label{fig5}
\end{figure}

\begin{figure}
\includegraphics[width=75mm]{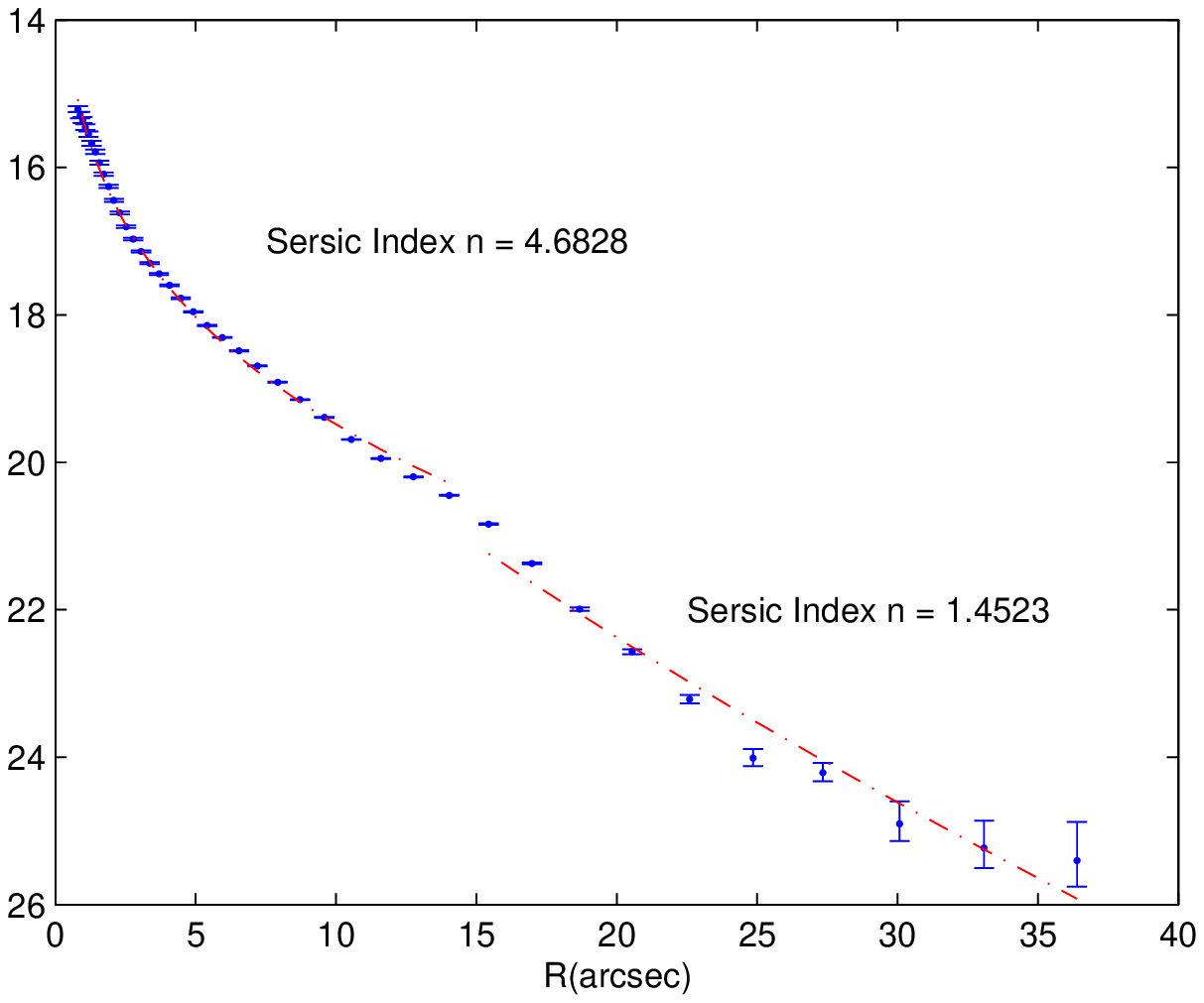}
\includegraphics[width=75mm]{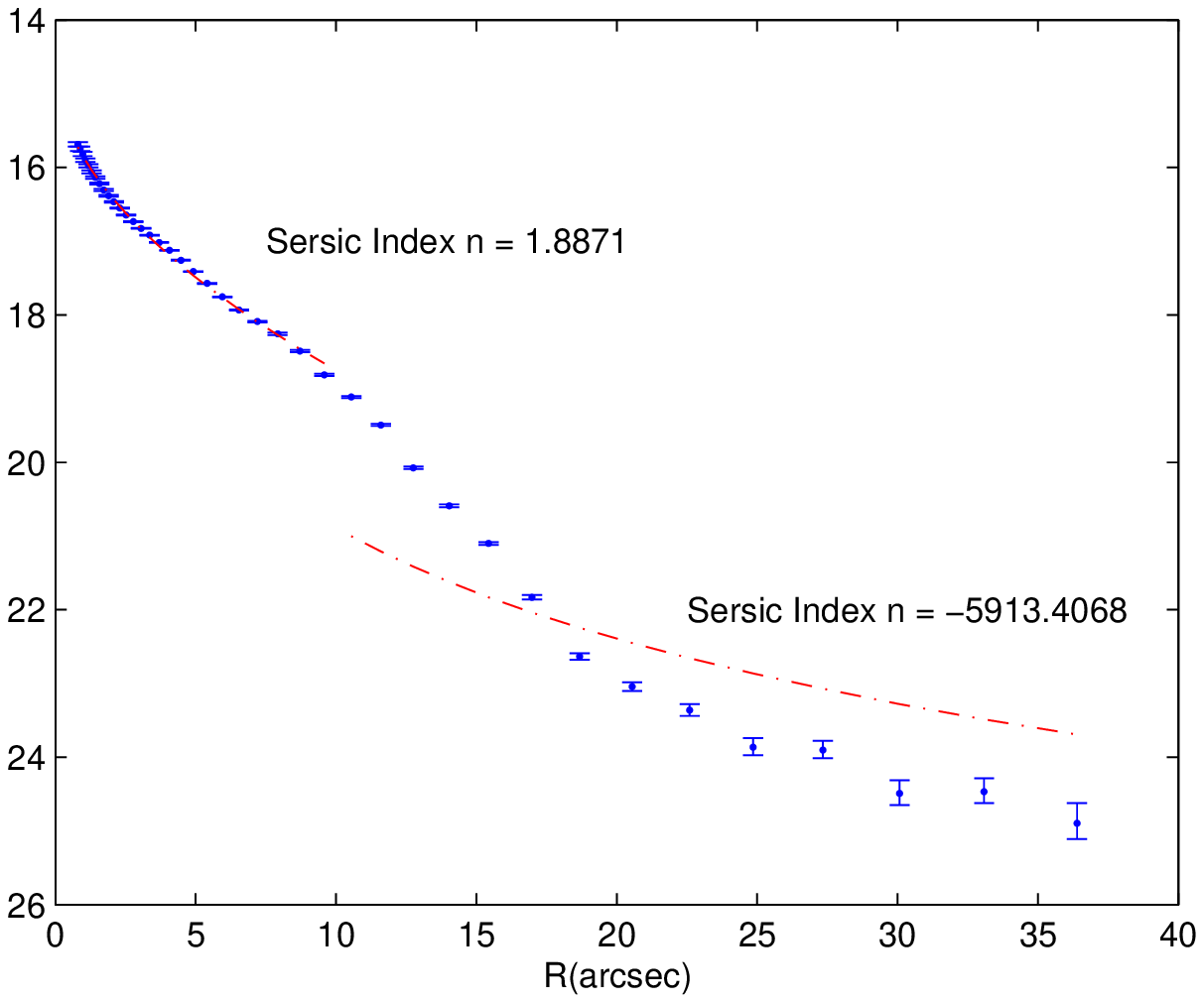}
\caption{Double S\'{e}rsic fits.The dashed line is the S\'{e}rsic's $R^{\frac{1}{n}}$ model. \quad Left: the LSB region reveals that it is actually an S0 system which may be misinterpreted as an elliptical without LSB region.\quad Right: double S\'{e}rsic profiles still fail to fit the curve, whose components can hardly be recognized by 1D S\'{e}rsic fit.}
   \label{fig6}
\end{figure}

In a word, we encounter the aforementioned four kinds of S\'{e}rsic fits and get a quite large range of S\'{e}rsic indices for the sample which is supposed to contain only strict early-types whose S\'{e}rsic indices should be closely concentrated to 4. The histogram of S\'{e}rsic indices presents the results clearly (Fig. 7 Left): the distribution of S\'{e}rsic indices has a peak at about 3-3.5 and a scatter from 1.5 to 8.5. The histogram of concentration indices shows a consistent result (Fig. 7 Right): most ETGs in our sample have concentration indices smaller than 2.5 while the typical value for ETGs
is $C_r=R_{\textrm{\scriptsize{90}}}/R_{\textrm{\scriptsize{50}}}>2.6$. These two histograms correlate with each other to imply that the LSB areas of the ETGs make a noticeable difference in S\'{e}rsic fits. On one hand, single S\'{e}rsic profile may be misleading and thus double-S\'{e}rsic fitting is needed. On the other hand, the LSB regions of the profiles make the ETGs' photometric properties shift to the disk end. 

\begin{figure}
\includegraphics[width=75mm]{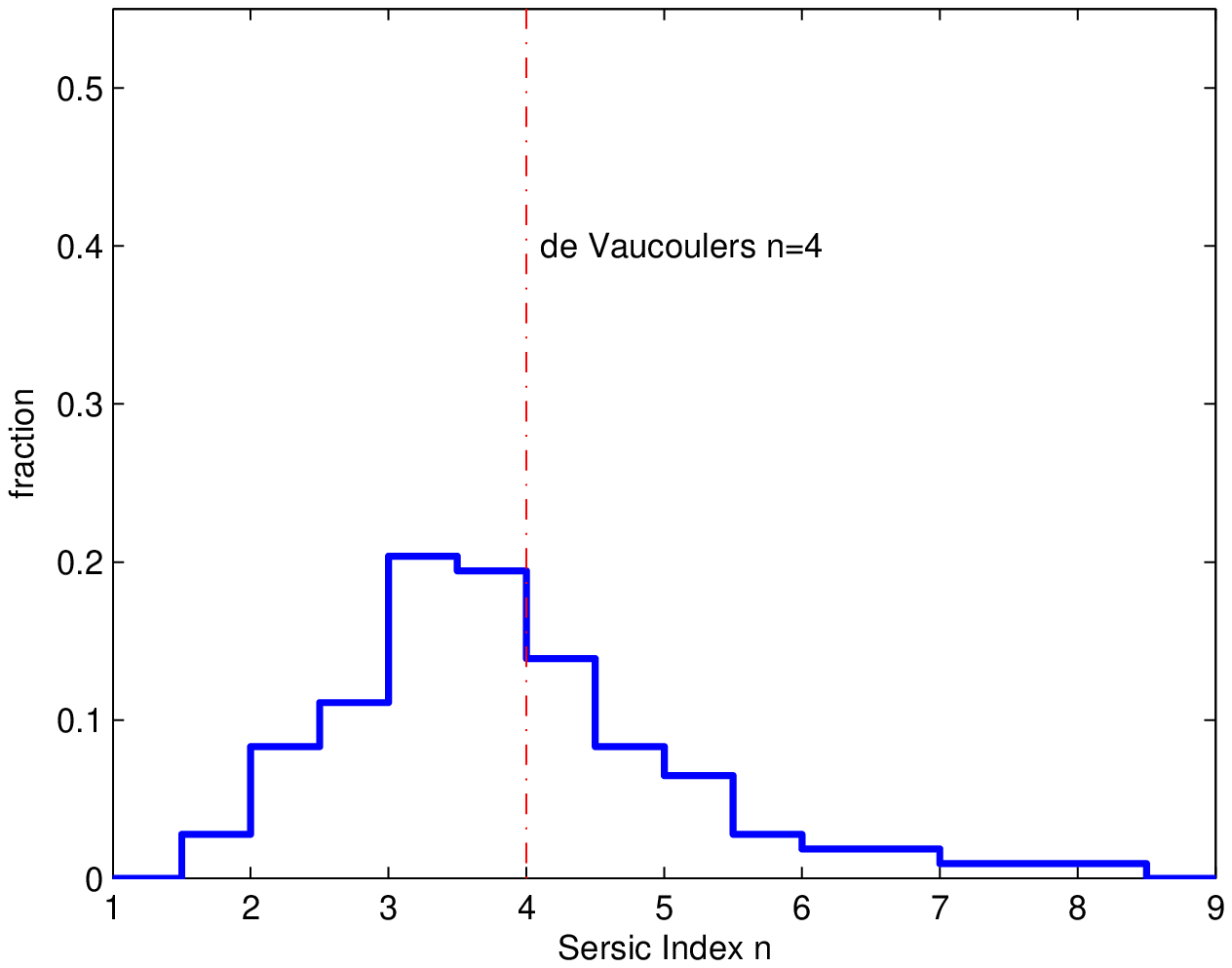}
\includegraphics[width=75mm]{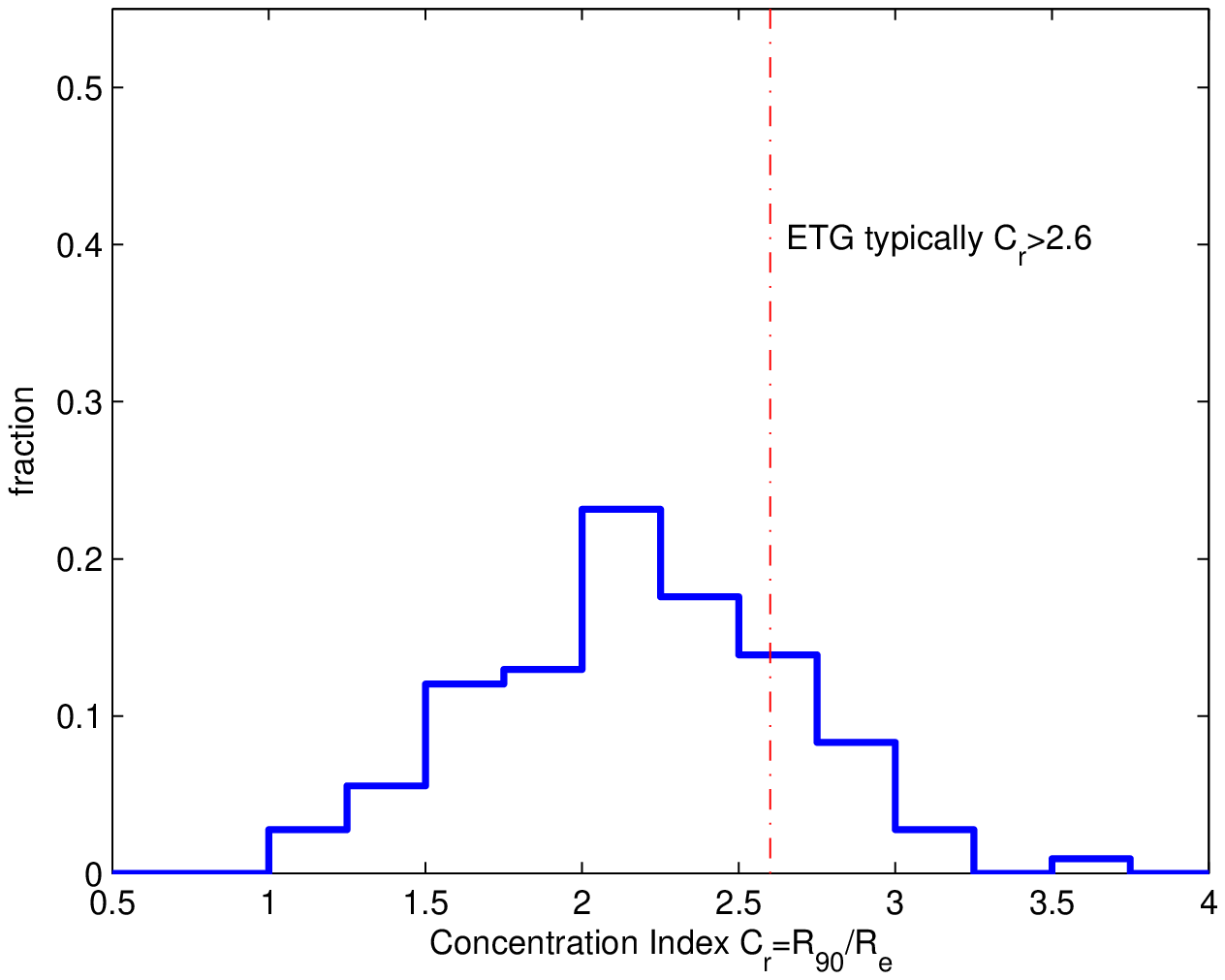}
\caption{Left: the histogram of S\'{e}rsic indices. The dashed line is where the peak ought to be for de Vaucouleur profiles. \quad Right: the histogram of concentration indices. The dashed line marks the position where the peak ought to be for typical ETGs. }
   \label{fig7}
\end{figure}

\subsection{Radial Color Profiles}
Radial color gradient is defined by its slope (Wu et al. 2005, Suh et al. 2010)
\begin{equation}
g_{g-r}\equiv \frac{d(g-r)}{d\log(\frac{R}{R_{\textrm{\scriptsize e}}})},\quad g_{u-r}\equiv \frac{d(u-r)}{d\log(\frac{R}{R_{\textrm{\scriptsize e}}})}.
\label{eq:color}
\end{equation}
We examine $g-r$ and $u-r$ profiles from 0.5$R_{\textrm{\footnotesize{e}}}$--1.5$R_{\textrm{\footnotesize{e}}}$ and 1.5$R_{\textrm{\footnotesize{e}}}$--4$R_{\textrm{\footnotesize{e}}}$ respectively in order to compare the slopes in the two colors and to compare the slopes in the two regions, paying extra attention to the outer LSB regions. The histograms of the $g-r$ and $u-r$ slopes tell us two direct statistical results (Fig. 8 Upper and Lower). Firstly, there are more positive gradients in LSB regions than in inner regions irrespective of $g-r$ or $u-r$. Secondly, there are more positive gradients in $g-r$ than in $u-r$ regardless of 0.5$R_{\textrm{\footnotesize{e}}}$--1.5$R_{\textrm{\footnotesize{e}}}$ or 1.5$R_{\textrm{\footnotesize{e}}}$--4$R_{\textrm{\footnotesize{e}}}$. 

\begin{figure}[h!!!]
\centering
\includegraphics[width=9.0cm, angle=0]{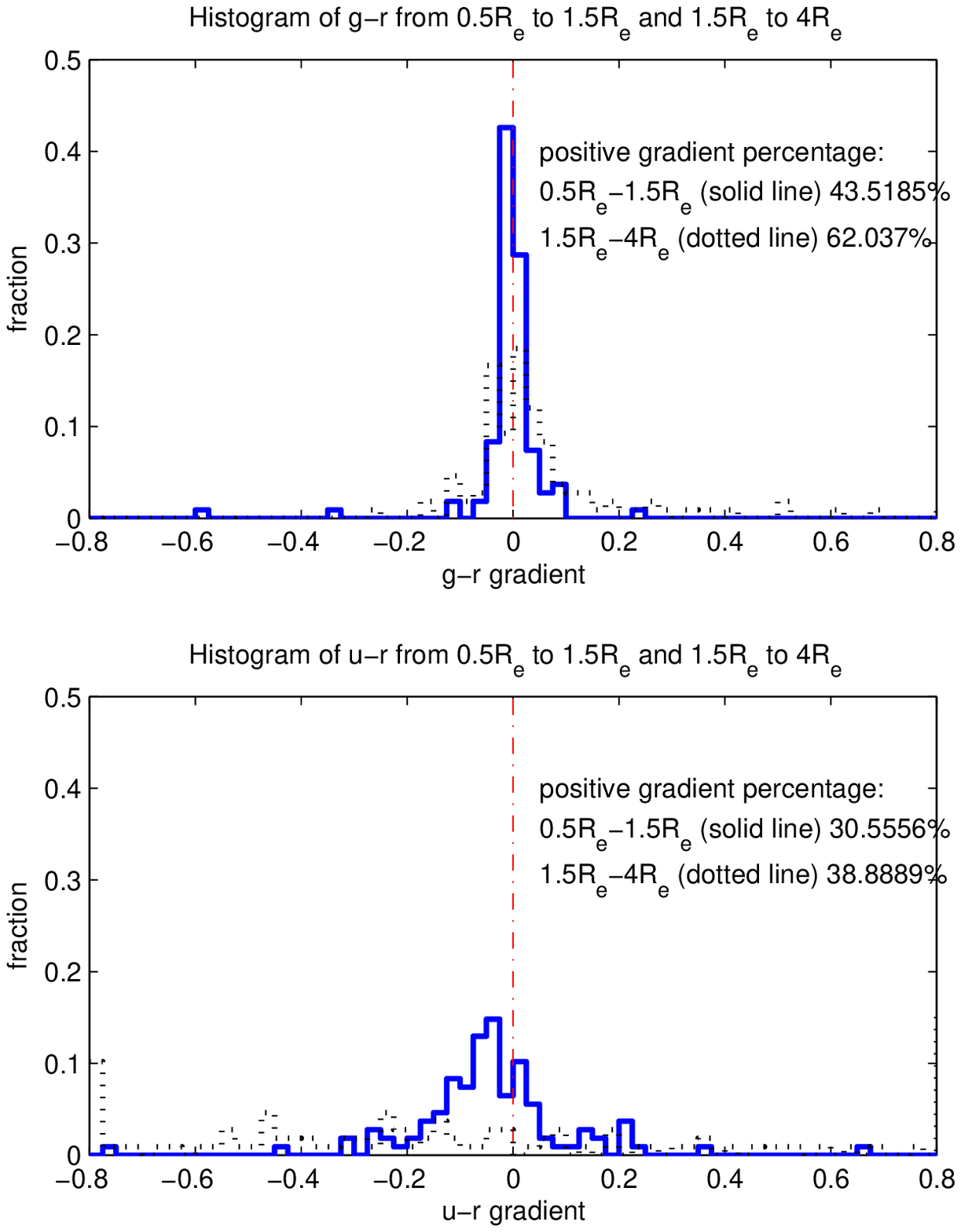}
\begin{minipage}[]{85mm}
\caption{ Upper: the histogram of $g-r$ gradients. \quad Lower: the histogram of $u-r$ gradients. \quad The solid lines are the histogram of the gradients in the inner region of 0.5$R_{\textrm{\footnotesize{e}}}$--1.5$R_{\textrm{\footnotesize{e}}}$, while the dotted lines are the histograms of the gradients in the outer 1.5$R_{\textrm{\footnotesize{e}}}$--4$R_{\textrm{\footnotesize{e}}}$ region. The dash-dot lines mark the threshold for positive/negative gradient.} 
\end{minipage}
\label{fig8}
\end{figure}

Now that we have two colors, we can define a 'red-core' ETG as follows: if both its $g-r$ and  $u-r$ color in the inner region show negative gradients, we call it a 'red-core' ETG. Similarly, we have a 'blue-core' galaxy if its $g-r$ and $u-r$ colors in the inner region both show positive gradients. According to these definitions, we have 62 red-core ETGs, taking up to 60\% of the whole sample and 13 blue-core ones, taking up to a much smaller portion, approximately 10\%. All the color gradients of all the 108 galaxies with reliable photometric results are listed in Appendix Table A.1 and Table A.2. The outcome that there are more red-core ETGs agree with the results of Suh et al. (2010), who report 11\% 'red-core' versus 4\% 'blue-core' in a large sample of 5002 ETGs, although the disagreement in the absolute portions has something to do with our different method of defining 'red-core'/'blue-core'. Since the blue-core ones are relatively rare, we would like to give an instance in Fig. 9, which may also help to illustrate how our color profiles look like.

\begin{figure}[h!!!]
\centering
\includegraphics[width=9.0cm, angle=0]{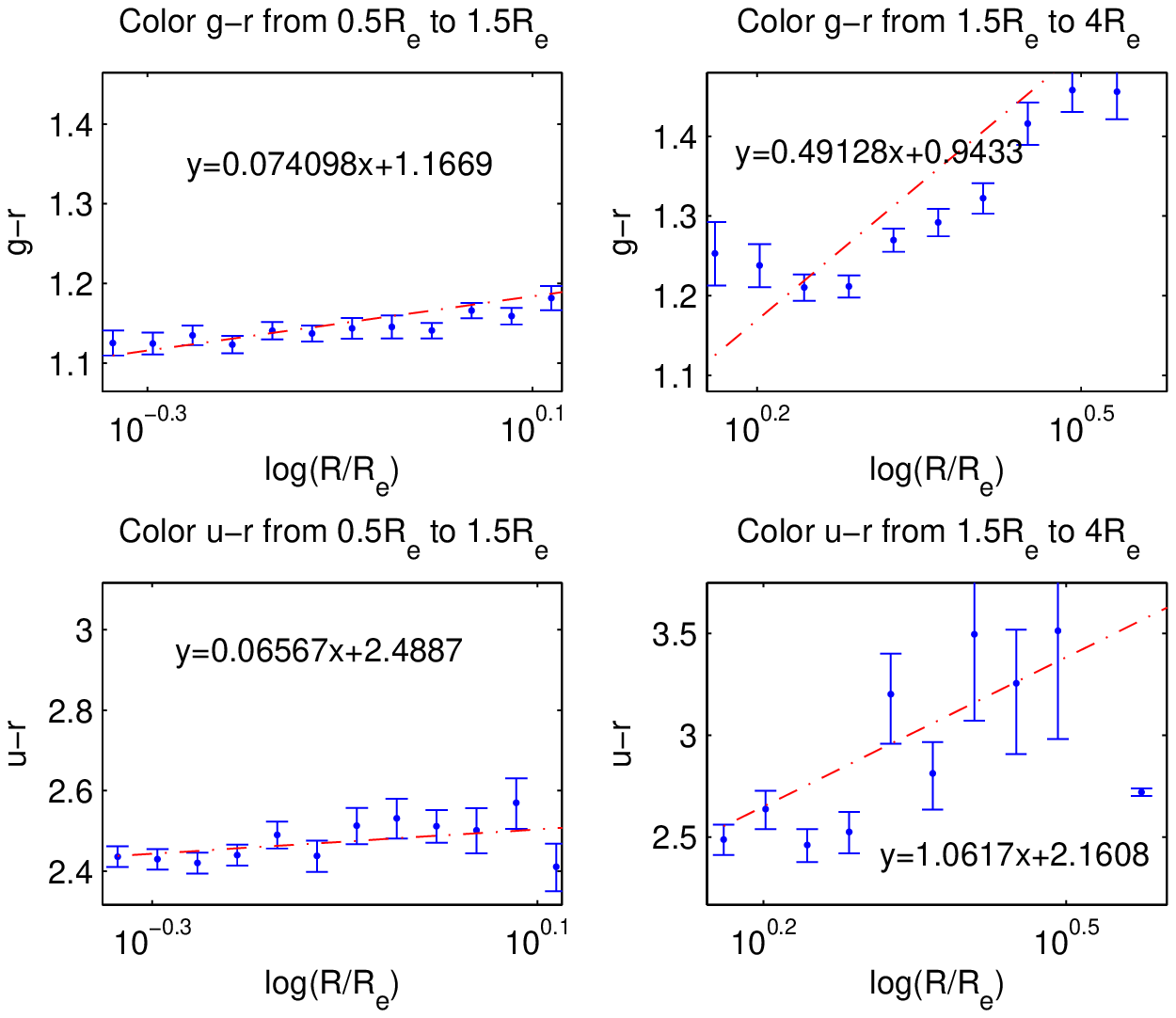}
\begin{minipage}[]{85mm}
\caption{ An example of the rare blue-core ETGs.} 
\end{minipage}
\label{fig9}
\end{figure}

When we move to the LSB regions, we find that 27 of the 62 red-core ETGs show positive $u-r$ gradient and 31 of the 62 red-core ETGs show positive $g-r$ gradient in their LSB regions respectively. 19 of the 62 red-core ETGs show both positive $u-r$ gradient and positive $g-r$ gradient in their LSB regions, taking up to almost 1/3 of all the red-core ETGs. Consistently, 10 of the 13 blue-core ETGs show positive $u-r$ gradient and 5 of the 13 blue-core ETGs show positive $u-r$ gradient in their LSB regions respectively. 4 of the 13 blue-core ETGs show both positive $u-r$ gradient and positive $g-r$ gradient in the LSB regions, also taking up to about 1/3 of the blue-core ETGs. All in all, the colors of red-core ETGs do not get bluer outwards monotonically, and the colors of blue-core ones do not get redder outwards monotonically either. Fig. 10 presents two examples of the 19 'abnormal' red-core ETGs. 

\begin{figure}
\includegraphics[width=75mm]{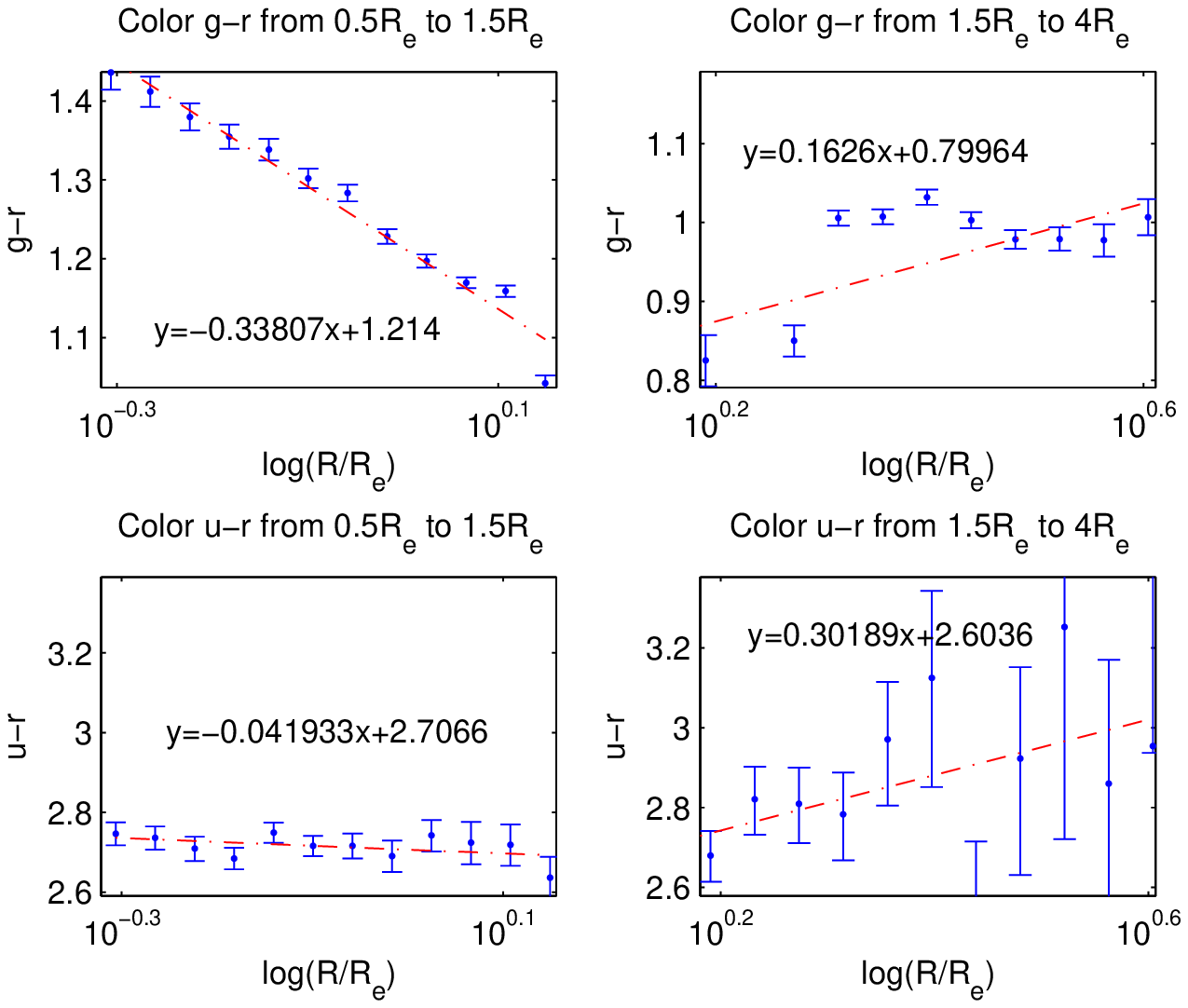}
\includegraphics[width=75mm]{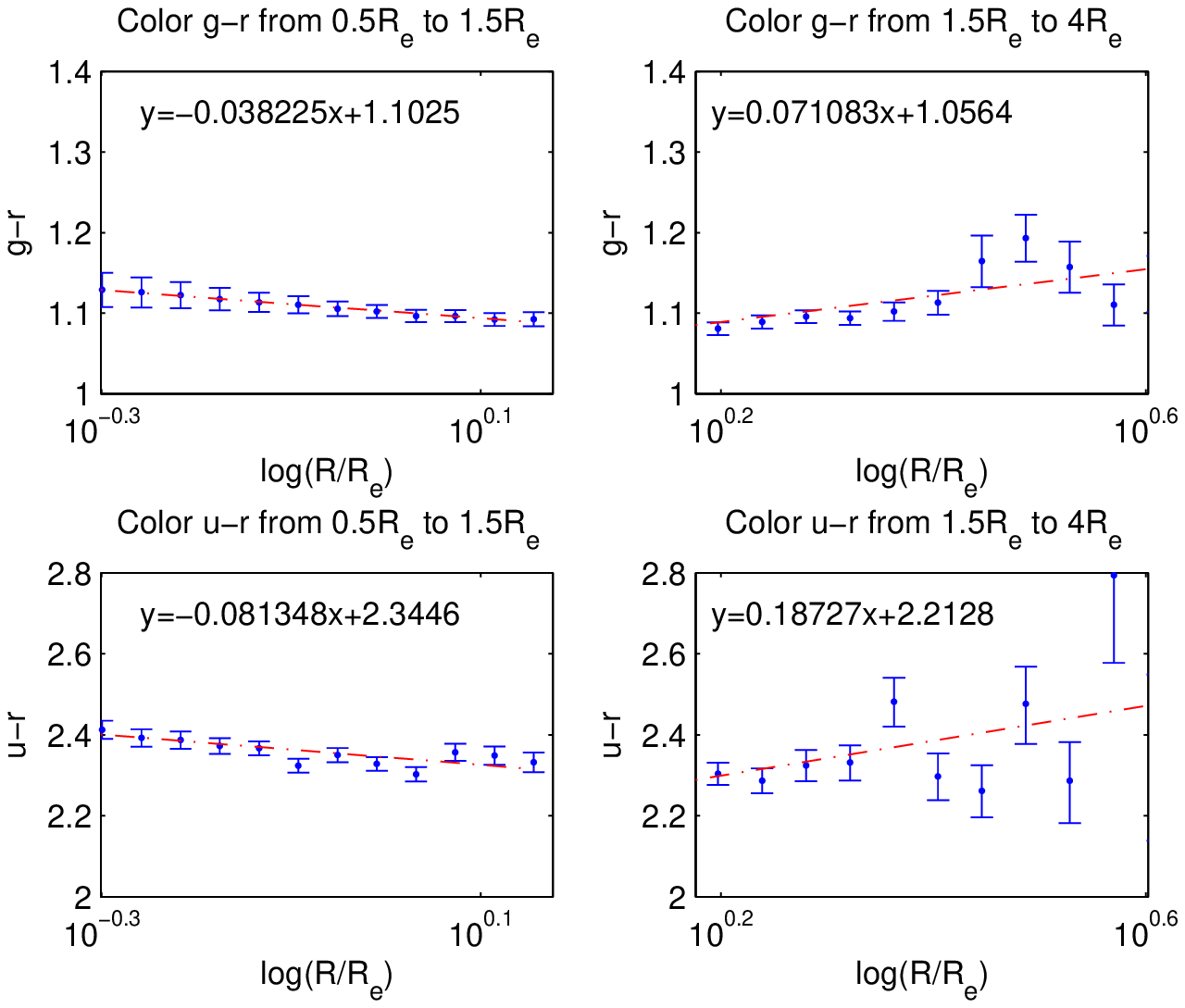}
\caption{Two examples of the red-core ETGs whose colors get redder, instead of bluer, beyond 1.5$R_{\textrm{\footnotesize{e}}}$. Red dashed lines indicate the slopes.}
   \label{fig10}
\end{figure}

\subsection{Isophotal Shapes}
ELLIPSE actually does not draw isophotes but a series of ellipses that approximately match the isophotes. The intensity along an ellipse is given in Fourier series:
\begin{equation}
I(\theta)=I_{\scriptsize 0}+\Sigma_{n=1}^{\infty}(A_n\cos n\theta +B_n\sin n\theta).
\label{eq:isophote}
\end{equation}
where $I_{\scriptsize 0}$ is the intensity averaged over the ellipse, and $A_n$, $B_n$ are the higher order Fourier coefficients. For isophotes of perfect ellipses, these coefficients should be zeros. 'A4', as the output of ELLIPSE, is actually $A_4$ divided by SMA $a$, i.e., $A_4/a$. Hao et al. (2006) proved in their appendix that $A_4/a$ is essentially just $a_4/a$, which is used by Bender et al. (1988, 1989) to tell whether an elliptical galaxy is disky ($a_4>0$) or boxy ($a_4<0$). $a_4/a$ is calculated from INTENS and A4. It is the weighted mean value of A4 over 2PSF to 1.5$R_{\textrm{\footnotesize{e}}}$ with intensity counts (INTENS) as the weight. $e$ is calculated similarly (Bender et al. 1988, 1989). From the histogram of $a_4/a$ (Fig. 11), we find that there are much more disky ETGs than boxy ones in our sample.

\begin{figure}[h!!!]
\centering
\includegraphics[width=6.0cm, angle=0]{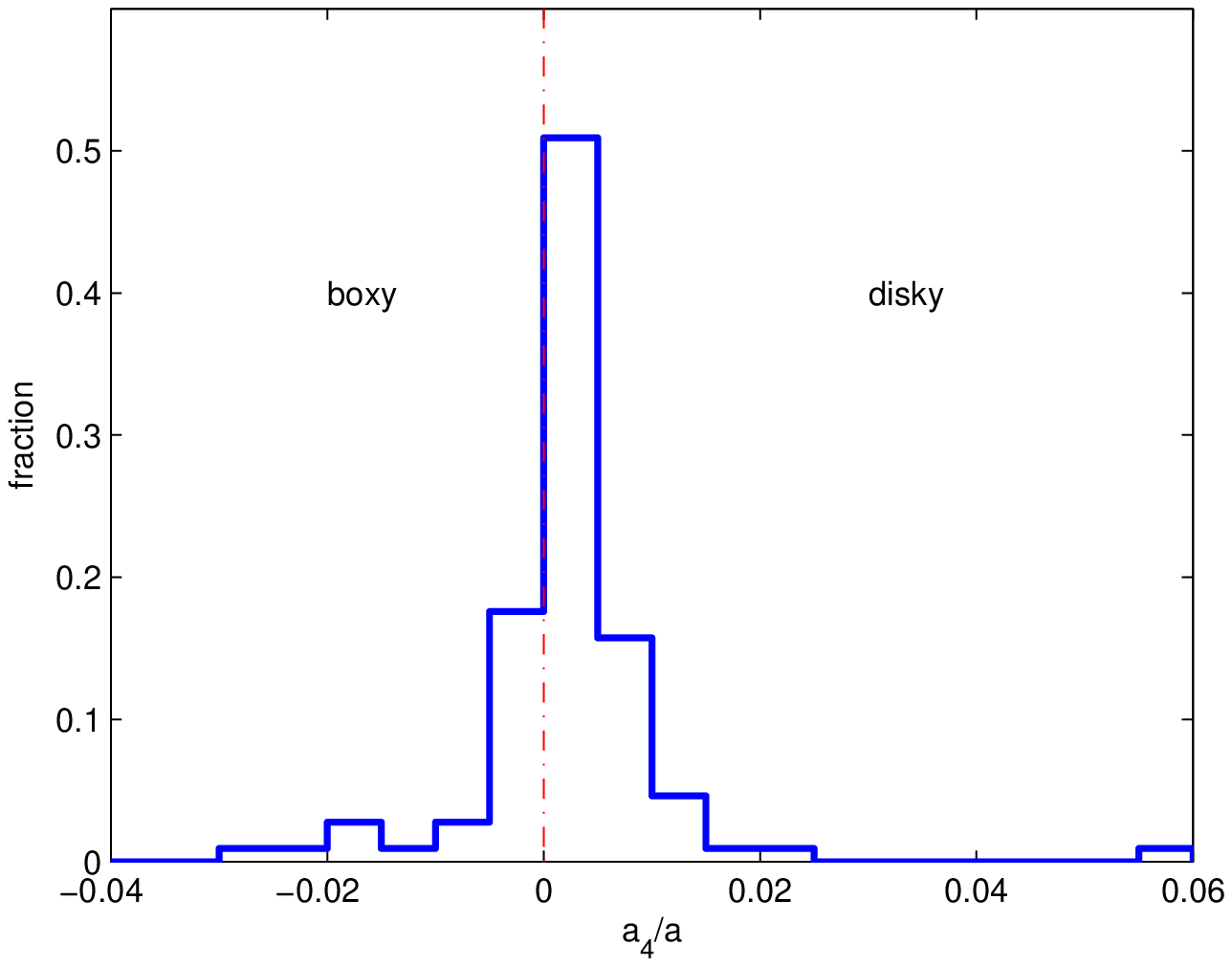}
\begin{minipage}[]{85mm}
\caption{ The histogram of $a_4/a$. The dash line is the threshold of disky/boxy, so there are much more disky ETGs in our sample.} 
\end{minipage}
\label{fig11}
\end{figure}

We plot $a_4/a$ versus $e$ in Fig. 12 to see whether or not at higher $e$, the deviations of the isophotes from perfect ellipses are larger (Hao et al. 2006). But due to the fact that our sample is not meant to be a complete and large one, we do not see clear trend in the diagram. However, it is plausible that at higher $e$, the ETGs are more likely to be S0s instead of Es, no wonder $a_4/a$ deviates from zero. 

\begin{figure}[h!!!]
\centering
\includegraphics[width=6.0cm, angle=0]{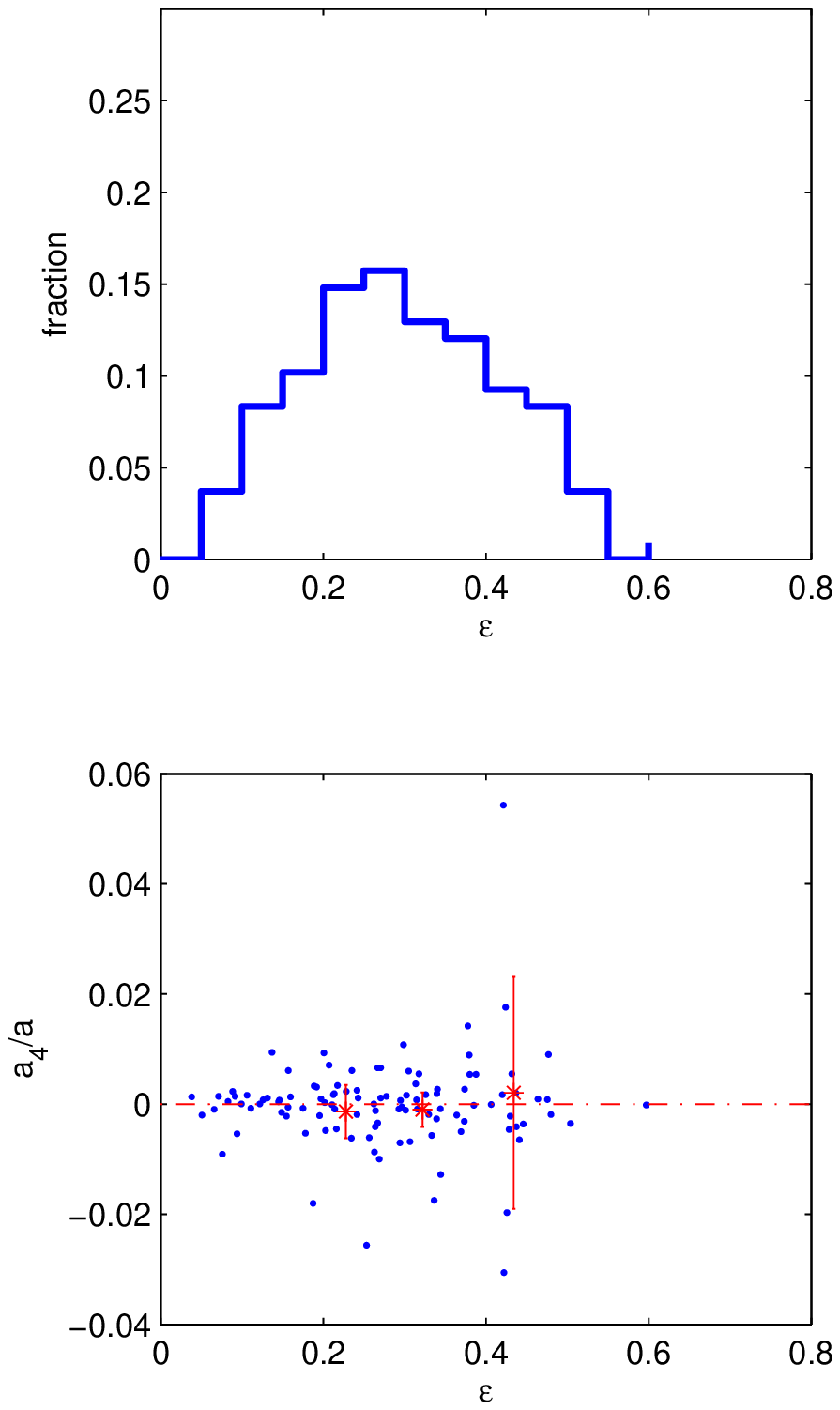}
\begin{minipage}[]{85mm}
\caption{ Upper: The histogram of $e$. \quad Lower: the $a_4/a$-$e$ diagram. The three stars and their error bars show the mean values of $a_4/a$ in each of the three bins and their scatter. The bins are divided according to the histogram of $e$ so that each bin contains approximately the same number of galaxies. We can see $a_4/a$ deviates from zero and its scatter increases as $e$ increases, although the trend is not convincing enough here. } 
\end{minipage}
\label{fig12}
\end{figure}

\section{Discussion}
\label{sect:discussion}

The results in Section 3.1 remind us that S\'{e}rsic fit may not be that robust as we thought before. In addition, the quality of S\'{e}rsic fitting relies much on where to truncate the surface brightness profiles. In a trial when we did not abandon the innermost 2PSF regions, we got the peak of S\'{e}rsic indices at around 2.5, the typical value for S0 systems. The innermost 2PSF regions are affected by seeing effects and possible stellar cusps. Truncating the outer region also matters: as we have tried double S\'{e}rsic profiles for the aforementioned two ETGs in Section 3.1, if we choose not to fit the outer regions purposely, the results of S\'{e}rsic fits may alter accordingly. Admittedly, this likelihood agrees with our discovery that the LSB regions matter a lot in 1D photometry and fitting.

Another issue is the definition of 'red-core'/'blue-core' ETGs. Suh et al. (2010) measured the $g-r$ color for 5002 galaxies and set a reliable threshold for 'red-core'/'blue-core', i.e., the $0.5\sigma$ confidence level (Figure 2., Suh et al. 2010). We think that this threshold works and helps a lot in picking out the ETGs with noticeable color gradient. However, we cannot use this definition in this work. Since we have a small sample of 111 luminous nearby galaxies, If we still stick to the threshold, then a large portion of our sample may fall into the category of 'no noticeable color gradient', leaving the number of the remaining ones not enough for statistical analyses. So we measure the $u-r$ color in addition to $g-r$ in the hope that the signs of the slopes of the two colors agree with each other, so that even if any single color cannot give a gradient above the threshold, the consistency between the two colors still approves a real color gradient. Fortunately, we examine all the ETGs to find that only one of them have a positive $g-r$ gradient and a negative $u-r$ slope. So we think that our definition of 'red-core'/'blue-core' is also reliable.

As to the fact that one third of red-core ETGs go red (one third of the blue-core ETGs go blue) in the LSB region, we think that a plausible interpretation involves galaxy mergers. Chang et al. (2006) found that massive ETGs are redder than less massive ones, and color is more sensitive to metallicity rather than age. Spolaor et al. (2010) point out that massive ETGs have less obvious metallicity gradients. These and other discoveries both imply that the opposite color gradients between the LSB regions and the inner regions may owe to metallicity gradients as remnant of mergers.

In the future, we can investigate the spectra of the LSB areas for the ETGs with opposite color gradients so as to resolve the age-metallicity degeneracy to see radial metallicity profiles.  For problems cannot be solved by 1D photometry, we can also try 2D surface photometry with tools such as GALFIT. 

\section{Summary}
\label{sect:summary}
In this paper, we present surface photometry and radial color gradients of nearby luminous early-type galaxies. The 2 magnitudes' extra depth of SDSS Stripe 82 provide us the opportunity to investigate into the low-surface-brightness areas. In the first place, we find that LSB regions make some S\'{e}rsic profiles of our high-FracDev ETGs deviate from de Vaucouleurs profile and shift to the exponential end. Then we find up to 60\% red-core ETGs and approximately 10\% blue-core ones, but red-core/blue-core ETGs do not necessarily have monotonic color gradients. On the contrary, about one third of the red-core/blue-core ETGs show opposite color gradients between inner regions and LSB regions, which may owe to galaxy mergers and subsequent radial metallicity gradients. Finally, we find that there are more disky ETGs in our sample and we also try to use the $a_4/a$-$e$ diagram to show that ETGs with higher $e$ values tend to have isophotes that deviate more from perfect ellipses.

\normalem
\begin{acknowledgements}
This work, as the Bachelor's thesis of Fangzhou Jiang, is supported by the National Fund for Fostering Talents of Basic Sciences of China. We thank all the members of the galactic and extragalactic group in the Department of Astronomy, Nanjing University for their helpful discussions. We gratefully acknowledges Lu Wei in the Department for Intensive Instruction of Nanjing University for her kind help with the data importation.
\end{acknowledgements}

\appendix                  
\section{Detailed information about the sample}
\begin{table}[h!!!]
\tiny
\centering
\begin{minipage}[]{120mm}
\caption[]{ Detailed Information about the Sample, Photometric Results and Color Gradients}\label{tab.a.1}\end{minipage}
\tabcolsep 0.2mm
 \begin{tabular}{cccccccccccccccc}
  \hline\noalign{\smallskip}
SDSS Name &  RA      & DEC & FracDev$_r$ & S\'{e}rsic Index & $R_{\textrm{\tiny{e}}}$ & $R_{\textrm{\tiny{90}}}$ & $\mu_{\textrm{\tiny{e}}}$ & $C_r$ & $g_{g-r, \textrm{\tiny{inner}}}$ & $g_{g-r, \textrm{\tiny{outer}}}$  & $g_{u-r, \textrm{\tiny{inner}}}$& $g_{u-r, \textrm{\tiny{outer}}}$ & $a_4/a$ & $e$ & $r$ Magnitudes  \\
  \hline\noalign{\smallskip}
J011248.6-001724.6	&18.20252444		&-0.29018160	&1	&3.5949	&9.4562	&19.33	&1.8851	&17.4809	&-0.018715	&-0.028358	&-0.07705	&-0.13112	&0.0017	&0.3259	&12.95\\
J235618.81-001820.1	&359.07837711	&-0.30560544	&1	&4.7676	&6.9561	&14.22	&1.6732	&17.3346	&-0.012698	&-0.0099012	&-0.095444	&-0.25043	&0.0037	&0.3138	&13.23\\
J010416.96-004553.6	&16.07067765	&-0.76491129	&0.89	&4.1974	&27.28	&30.67	&1.293	&19.2839	&-0.010488	&-0.091517	&0.117		&0.23739		&-0.00090516	&0.3156	&13.27\\
J011146.55-003951.5	&17.94396288	&-0.66432211	&0.97	&2.6587	&4.1057	&10.31	&2.3159	&16.2258	&0.045424	&-0.010903	&0.014372	&-0.12744	&0.0019	&0.2135	&13.37\\
J012943.98-011429.1	&22.43327416	&-1.24141730	&0.93	&2.8856	&11.045	&21.7	&1.9771	&17.816	&0.0047963	&0.044714	&-0.0092825	&-0.15645	&-0.000063854	&0.4064	&13.38\\
J030732.29-005752.3	&46.88455434	&-0.96453808	&0.97	&3.225	&9.4557	&21.1	&2.2342	&17.7391	&-0.016407	&0.060442	&-0.12452	&-0.28238	&0.0034	&0.2176	&13.38\\
J015441.04-000836		&28.67100555	&-0.14334793	&0.98	&4.0803	&11.894	&23.82	&1.9683	&18.2665	&-0.023882	&-0.093205	&-0.042127	&-0.23902	&0.0066	&0.2705	&13.39\\
J030530.84-002418		&46.37852752	&-0.40501239	&1	&3.0552	&8.534	&17.64	&1.9687	&17.5945	&-0.017979	&0.019746	&-0.079311	&0.25322	&0.0019	&0.3397	&13.42\\
J231808.39-002325.7	&349.53496202	&-0.39047265	&0.8	&5.3163	&4.665	&20.69	&3.5383	&19.0252	&-0.031405	&-0.13732	&0.0093273	&0.75419	&0.0013	&0.0382	&13.5\\
J011853.62-010007.2	&19.72342956	&-1.00200616	&0.88	&3.6808	&20.9526	&30.88	&1.7327	&19.0464	&-0.021079	&0.25975	&0.059285	&-0.019325	&-0.0057	&0.3335	&13.53\\
J230748.93+005625.9	&346.95391663	&0.94053502	&1	&4.0026	&8.0254	&16.13	&1.8922	&17.8216	&-0.014986	&-0.039578	&-0.047298	&0.085562	&-0.00077092	&0.175	&13.6\\
J232029.07-010008.7	&350.12112518	&-1.00242678	&1	&4.5442	&8.8691	&18.31	&2.027	&18.1781	&-0.018996	&0.045736	&-0.16763	&-0.44829	&-0.002	&0.0508	&13.68\\
J012602.51-011332.1	&21.51047343	&-1.22560582	&0.96	&3.9325	&12.812	&26.75	&2.2357	&18.6633	&0.0041238	&0.093578	&-0.05713	&-0.1852	&-0.018	&0.1875	&13.72\\
J011457.58+002550.9	&18.73995395	&0.43082652	&0.79	&5.3822	&33.3274	&23.91	&0.99779	&19.7342	&-0.025616	&0.016946	&-0.2406	&-0.50457	&-0.0061	&0.2565	&13.76\\
J012312.36-003828.2	&20.80153809	&-0.64116891	&0.98	&2.6343	&8.8963	&14.57	&1.5459	&17.6697	&-0.029947	&-0.017055	&-0.12331	&-0.33671	&-0.00015011	&0.5972	&13.8\\
J234430.08+001913.7	&356.12533533	&0.32047532	&0.97	&3.6178	&8.1598	&22	&2.6638	&18.1101	&-0.010702	&-0.0081651	&-0.0052486	&0.57781	&0.0023	&0.0888	&13.81\\
J031212.86-010342.8	&48.05360848	&-1.06190079	&0.98	&3.1603	&7.2191	&15.46	&2.0999	&17.75	&-0.038225	&0.071083	&-0.081348	&0.18727	&-0.0041	&0.2641	&13.94\\
J233525.66+010332.6	&353.85695746	&1.05907480	&0.93	&3.492	&15.831	&26.16	&1.8106	&19.1562	&-0.0091442	&-0.18197	&-0.16693	&0.63191	&-0.0256	&0.2533	&13.97\\
J001647+004215.3		&4.19586472	&0.70427291	&0.95	&3.5331	&10.1844	&19.4	&1.896	&18.5757	&-0.029121	&-0.040562	&-0.09992	&-0.34979	&0.0025	&0.2417	&14.03\\
J222614.63+004004	&336.56097640	&0.66778611	&1	&3.1711	&6.5502	&16.01	&2.4751	&17.6089	&0.024994	&0.0093977	&0.13477	&-0.15963	&-0.0027	&0.3393	&14.06\\
J220425.31+004255.5	&331.10546741	&0.71542132	&1	&2.5506	&5.4687	&13.01	&2.181	&17.309	&-0.028337	&-0.035825	&-0.13994	&-0.43243	&0.0017	&0.4201	&14.11\\
J204710.5+002147.8	&311.79378369	&0.36329981	&0.81	&2.0488	&4.3964	&11.11	&2.3169	&16.9985	&-0.11481	&-0.071288	&-0.26335	&1.5204	&0.0011	&0.243	&14.11\\
J021735.8-002936.8	&34.39917627	&-0.49357706	&0.98	&3.3831	&6.7671	&15.73	&2.1545	&18.1969	&-0.017447	&-0.029389	&-0.77129	&-0.22874	&0.0014	&0.0713	&14.15\\
J033601.58+010617.1	&54.00660523	&1.10475887	&1	&4.0362	&8.5412	&14.56	&1.601	&18.3207	&-0.01871	&0.025516	&-0.047585	&1.1535	&0.0027	&0.3738	&14.16\\
J020557.03+004624	&31.48762819	&0.77334047	&0.98	&4.4026	&10.4865	&18.4	&1.7924	&18.731	&-0.016095	&0.037535	&-0.052994	&-0.39026	&0.0093	&0.2007	&14.16\\
J205838.24+005445.3	&314.65936086	&0.91258449	&0.99	&4.0209	&10.3188	&21.31	&2.0919	&18.8015	&-0.021973	&0.069314	&-0.052994	&0.46902	&0.0094	&0.1371	&14.16\\
J231110.01-000908		&347.79174779	&-0.15224910	&1	&3.0104	&5.1929	&15.15	&2.6967	&17.6412	&0.00028358	&0.011971	&-0.084359	&-0.016687	&-0.0022	&0.1547	&14.19\\
J024647.79-003414.8	&41.69915516	&-0.57079231	&0.95	&5.0163	&18.496	&28.24	&1.9574	&19.4473	&-0.031991	&0.019586	&-0.0099847	&0.9781	&0.000035055	&0.1219	&14.19\\
J233318.7-002700.5	&353.32792549	&-0.45014581	&1	&4.5104	&20.0121	&25.79	&1.5124	&19.6328	&-0.038791	&-0.055995	&0.14284	&-0.78908	&0.00077529	&0.3147	&14.19\\
J005727.62-002817.6	&14.36509352	&-0.47157061	&1	&8.1125	&14.793	&12	&0.90306	&19.0577	&-0.012473	&-0.048233	&-0.26462	&-0.46491	&-0.005	&0.3693	&14.2\\
J014017.46-001740.8	&25.07278726	&-0.29467689	&1	&3.9869	&9.6875	&19.01	&2.1259	&18.4792	&-0.008886	&0.60998	&0.64622	&1.6074	&-0.0062	&0.2347	&14.21\\
J222904.69-011105.6	&337.26956145	&-1.18490971	&0.91	&2.6933	&10.5183	&19.23	&1.8602	&18.5058	&0.010871	&0.10962	&-0.0038694	&-0.76463	&-0.0046	&0.4284	&14.21\\
J212413.05+010706.1	&321.05439566	&1.11838228	&1	&4.4661	&9.0369	&12.61	&1.4091	&18.2644	&-0.023212	&0.014296	&-0.025948	&0.31981	&0.0054	&0.3878	&14.22\\
J002537.7+000212.2	&6.40709052	&0.03672977	&1	&3.5388	&6.0791	&13.46	&1.9563	&17.8479	&-0.065434	&0.013746	&0.031839	&1.0394	&-0.0128	&0.3443	&14.22\\
J005130.06-011511.1	&12.87528325	&-1.25309676	&0.93	&6.4237	&10.0964	&16.75	&1.3652	&19.0911	&0.035644	&0.67256	&-0.203	&-0.82407	&-0.0019	&0.4801	&14.25\\
J015113.38-010338.5	&27.80576264	&-1.06070791	&0.93	&7.7085	&5.9679	&11.92	&1.1762	&18.862	&0.0078687	&0.10491	&-0.093857	&-0.90778	&0.0011	&0.2707	&14.26\\
J222812.65+003235.6	&337.05272498	&0.54322348	&0.95	&3.8585	&10.6137	&21.27	&2.0915	&18.9029	&-0.042947	&-0.031789	&-0.23347	&-0.55431	&-0.0054	&0.0941	&14.26\\
J034313.96-010107.3	&55.80819172	&-1.01869778	&0.88	&2.8164	&7.439	&16.64	&2.1299	&18.3023	&-0.039968	&0.083739	&-0.093829	&0.029362	&-0.0053	&0.178	&14.26\\
J012235.4-003412.3	&20.64751590	&-0.57009991	&1	&3.5454	&5.2537	&15	&3.0277	&17.3873	&-0.024095	&0.0089256	&-0.063819	&0.66851	&-0.0034	&0.267	&14.26\\
J010425.27-001114.6	&16.10529594	&-0.18740242	&0.97	&3.1162	&11.872	&20.54	&1.7197	&18.8413	&-0.0081743	&0.076599	&0.018154	&0.49289	&0.002	&0.4369	&14.27\\
J022026.53-001846.2	&35.11057621	&-0.31283885	&0.95	&2.9008	&9.3235	&14.55	&1.623	&18.2389	&-0.0090067	&-0.010362	&-0.10478	&-0.28946	&0.00092875	&0.4641	&14.29\\
J032516.26-002432.4	&51.31775271	&-0.40902744	&1	&2.8124	&8.1868	&17.12	&2.0769	&18.3298	&0.027014	&-0.0033398	&0.021128	&-0.084288	&0.009	&0.477	&14.29\\
J224005.22-004827.7	&340.02175136	&-0.80771312	&1	&3.3358	&5.9875	&12.03	&1.769	&17.9762	&-0.33807	&0.1626	&-0.041933	&0.30189	&-0.00091017	&0.293	&14.32\\
J012802.13-004417.9	&22.00888578	&-0.73832185	&0.92	&2.945	&6.5298	&13.81	&2.0069	&18.1141	&0.00088229	&-0.011681	&0.029804	&0.18113	&0.0066	&0.2671	&14.33\\
J020015.67+001157.2	&30.06529850	&0.19923396	&0.9	&1.9656	&5.8112	&14.71	&2.5827	&17.6397	&-0.012259	&-0.04439	&-0.032495	&0.81589	&0.006	&0.3049	&14.35\\
J012117.6-003551.5	&20.32335317	&-0.59764847	&1	&3.5104	&5.5602	&12.64	&1.9067	&17.9592	&-0.015385	&0.036747	&-0.06496	&0.16729	&0.0055	&0.3177	&14.38\\
J030916.97-005424		&47.32072761	&-0.90666964	&1	&5.955	&6.397	&11.87	&1.3835	&18.6907	&-0.019572	&0.26896	&-0.1261	&-0.48362	&-0.0045	&0.2162	&14.41\\
J005716.97-004010		&14.32072141	&-0.66947146	&0.92	&1.7408	&6.2232	&17.06	&2.8976	&17.8749	&0.0022425	&-0.049961	&0.0064705	&-0.24768	&0.0011	&0.1315	&14.48\\
J233941.25-000204.1	&354.92188685	&-0.03447809	&1	&2.3442	&5.8469	&14.17	&2.456	&17.7817	&-0.02072	&-0.011135	&-0.13553	&-0.23719	&0.00069408	&0.295	&14.49\\
J212258.68+010136.1	&320.74452146	&1.02670285	&0.97	&1.8637	&4.0267	&10.39	&2.5369	&16.9602	&-0.018559	&-0.06997	&-0.099956	&-0.65464	&-0.0019	&0.3298	&14.49\\
J235748.57+004744.1	&359.45240705	&0.79558976	&1	&4.7676	&8.843	&14	&1.5217	&18.7963	&0.034554	&0.034517	&0.002356	&-0.48148	&-0.00011651	&0.211	&14.49\\
J005610.66-010700.8	&14.04444799	&-1.11690788	&0.97	&2.2418	&6.1184	&16.5	&2.6009	&17.9435	&-0.016162	&0.49838	&-0.035228	&1.0623	&0.0027	&0.3403	&14.5\\
J002032.94-000847	 $^a$	&5.13726016	&-0.14639341	&1		&--- ---	&--- ---	&--- ---	&--- ---	&--- ---	&--- ---	&--- ---	&--- ---	&--- ---	&--- ---	&--- ---	&--- ---\\
J013503.43-005427.6	&23.76430097	&-0.90767569	&1	&5.8844	&18.477	&21.66	&1.5316	&19.597	&-0.04285	&0.391	&-0.046522	&0.45009	&-0.0175	&0.3364	&14.5\\
J013314.93-003813		&23.31221838	&-0.63695538	&1	&3.7814	&10.9742	&21.43	&2.1469	&18.9459	&0.00021205	&-0.058668	&-0.15977	&-0.22324	&-0.01	&0.2691	&14.51\\
J034745.73+004112.2 $^a$	&56.94055494	&0.68673663	&1	&--- ---	&--- ---	&--- ---	&--- ---	&--- ---	&--- ---	&--- ---	&--- ---	&--- ---	&--- --- 	&--- ---	&--- ---\\
J001312.19+004434.6	&3.30083192	&0.74294446	&1	&4.7469	&5.8391	&13.66	&2.0103	&18.4075	&-0.050437	&0.016114	&-0.1478	&-0.72481	&0.0023	&0.2281	&14.51\\
J012635.01-010646.7	&21.64587872	&-1.11299195	&1	&2.6173	&7.1553	&18.09	&2.5561	&18.2007	&0.013971	&0.11967	&0.054545	&-0.027389	&0.0176	&0.4243	&14.51\\
J001646.29+011237.1	&4.19288616	&1.21031392	&1	&3.8932	&6.883	&16.36	&2.156	&18.5188	&-0.023887	&-0.05461	&-0.07255	&-0.089308	&0.0017	&0.2124	&14.53\\
  \noalign{\smallskip}\hline
\end{tabular}
\tablenotes{a}{0.95\textwidth}{The photometric results for J002032.94-000847 and J034745.73+004112.2 are not reliable, so it is not taken into the analyses afterwards.}
\end{table}

\begin{table}[h!!!]
\tiny
\centering
\begin{minipage}[]{120mm}
\caption[]{ (Continue) Detailed Information about the Sample, Photometric Results and Color Gradients}\label{tab.a.2}\end{minipage}
\tabcolsep 0.2mm
 \begin{tabular}{cccccccccccccccc}
  \hline\noalign{\smallskip}
SDSS Name &  RA      & DEC & FracDev$_r$ & S\'{e}rsic Index & $R_{\textrm{\tiny{e}}}$ & $R_{\textrm{\tiny{90}}}$ & $\mu_{\textrm{\tiny{e}}}$ & $C_r$ & $g_{g-r, \textrm{\tiny{inner}}}$ & $g_{g-r, \textrm{\tiny{outer}}}$  & $g_{u-r, \textrm{\tiny{inner}}}$& $g_{u-r, \textrm{\tiny{outer}}}$ & $a_4/a$ & $e$ & $r$ Magnitudes  \\
  \hline\noalign{\smallskip}
J002339.87-000715		&5.91616577	&-0.12084572	&0.92	&2.9463	&11.7017	&21.52	&1.884	&19.2318	&-0.0089079	&-0.037576	&0.014278	&-0.54903	&-0.0035	&0.5041	&14.53\\
J002900.99-011341.7	&7.25412553	&-1.22825889	&1	&3.6191	&9.5859	&14.24	&1.5168	&18.686	&0.22002	&-0.12594	&0.19745	&-0.75826	&-0.0306	&0.4221	&14.54\\
J213645.79+011456.3	&324.19080001	&1.24899874	&1	&4.8758	&5.8036	&13.33	&2.0949	&18.1871	&-0.021301	&0.037401	&-0.00039655	&1.1429	&0.00071871	&0.146	&14.54\\
J024940.69-003357.2	&42.41956744	&-0.56591588	&0.97	&2.5231	&5.5643	&14.16	&2.4173	&17.9421	&-0.029371	&0.00039767	&-0.0728	&0.35647	&0.000049035	&0.2622	&14.54\\
J001642.55-002643.5	&4.17729552	&-0.44542089	&1	&4.9808	&8.2133	&15.42	&1.969	&18.6272	&-0.012721	&0.044825	&-0.066157	&-0.39334	&0.0007837	&0.1259	&14.56\\
J015315.24+010220.7	&28.31350548	&1.03908428	&0.78	&3.4584	&24.3583	&24.68	&1.1521	&20.3325	&-0.01987	&-0.03403	&-0.063034	&-0.051731	&-0.0068	&0.3068	&14.57\\
J011913.49-010839.9	&19.80624066	&-1.14442738	&0.96	&1.9575	&6.4876	&16.19	&2.5468	&17.999	&-0.03556	&-0.01556	&-0.057732	&-0.25693	&-0.00087108	&0.3441	&14.6\\
J013132.91+003321.5	&22.88714103	&0.55598754	&1	&5.4778	&39.3403	&23.77	&1.0447	&20.3068	&-0.01927	&-0.27174	&-0.14258	&-1.1361	&0.0543	&0.4216	&14.6\\
J002819.3-001446.8	&7.08043351	&-0.24633481	&1	&3.3719	&5.236	&13.49	&2.1941	&18.0847	&-0.6122	&0.01018	&-0.042758	&-0.37971	&-0.002	&0.3641	&14.64\\
J013127.62+010947	&22.86509079	&1.16307996	&0.96	&3.5394	&11.4009	&19.92	&1.8162	&19.2226	&0.071661	&-0.13638	&-0.029998	&-0.79335	&0.00083218	&0.4757	&14.64\\
J020316.02-010225.1	&30.81675458	&-1.04031200	&1	&3.9051	&7.2419	&15.62	&1.8976	&18.819	&-0.0072916	&0.14995	&-0.25452	&-1.2654	&0.0033	&0.1889	&14.65\\
J213500.39-003041.1	&323.75162529	&-0.51143543	&0.97	&2.9277	&8.1004	&19.84	&2.5201	&18.6855	&-0.03294	&0.034419	&-0.10639	&0.053361	&-0.0087	&0.2628	&14.67\\
J002949.34-011405.7	&7.45561756	&-1.23492103	&1	&7.0024	&12.3966	&23.66	&2.6859	&18.8819	&0.074098	&0.49128	&0.06567	&1.0617	&0.0061	&0.2352	&14.68\\
J223924.03-010216		&339.85014609	&-1.03778562	&1	&3.8234	&6.1246	&16.1	&2.339	&18.477	&0.0021905	&0.16388	&-0.12946	&-0.67801	&-0.0048	&0.2028	&14.68\\
J014715.74+005748.1	&26.81560871	&0.96336361	&0.93	&1.9943	&5.2313	&16.05	&3.023	&17.993	&-0.027968	&-0.011202	&0.12215	&0.40489	&0.0013	&0.1596	&14.7\\
J024513.77-004446		&41.30737872	&-0.74613175	&1	&5.2114	&10.7612	&15.13	&1.5141	&19.1134	&-0.017336	&-0.054555	&-0.12124	&-0.45631	&0.0108	&0.2986	&14.7\\
J011454.25+001811.8	&18.72605022	&0.30327972	&0.88	&2.3265	&5.0339	&12.92	&2.4188	&17.9362	&-0.038181	&-0.062553	&-0.081268	&1.1149	&0.00029361	&0.2019	&14.71\\
J234456.11+010736.8	&356.23380146	&1.12691012	&0.93	&2.33	&6.9886	&18.96	&2.6965	&18.5893	&-0.06396	&0.25715	&-0.13066	&-0.18992	&0.0061	&0.157	&14.73\\
J234107.25+000542.8	&355.28021641	&0.09524386	&1	&3.3672	&9.3763	&16.99	&1.8217	&19.0061	&-0.0078595	&0.00049809	&0.19783	&1.3148	&-0.0011	&0.3014	&14.74\\
J011204.62-001442.3	&18.01925392	&-0.24511037	&0.97	&4.79	&27.7696	&25.74	&1.5459	&20.0675	&-0.0045894	&-0.11305	&0.20229	&-2.1856	&-0.0041	&0.4376	&14.75\\
J024716.96-002325.2	&41.82067259	&-0.39034732	&1	&3.2473	&5.9714	&14.47	&2.3837	&18.3683	&0.00025596	&-0.040222	&-0.0017929	&0.97321	&-0.00055588	&0.1566	&14.75\\
J015518.65+002912.3	&28.82771889	&0.48675415	&1	&3.0868	&4.1656	&12.07	&2.6577	&17.7805	&0.0044897	&0.2277	&0.097484	&0.072748	&0.0014	&0.0917	&14.76\\
J235545.2-011519.6	&358.93833991	&-1.25544995	&0.99	&2.6916	&5.4229	&14.53	&2.7658	&17.9402	&0.0013852	&0.062428	&-0.0554438	&-0.29238	&-0.00091716	&0.2142	&14.76\\
J220638.46+011036.8	&331.66025782	&1.17690392	&1	&2.9806	&5.1769	&12.51	&2.1242	&18.1196	&-0.019533	&-0.048991	&-0.081213	&-0.42923	&0.0016	&0.3023	&14.77\\
J003933.21+003550.9	&9.88839858	&0.59748157	&0.95	&4.3542	&11.27	&16.04	&1.3079	&19.3581	&-0.0049445	&-0.0079676	&-0.20375	&-0.53938	&-0.0065	&0.4412	&14.77\\
J032833.19+010024.2	&52.13830364	&1.00672494	&1	&6.5125	&20.8425	&20.67	&1.1421	&20.3533	&-0.04183	&-0.058231	&-0.44186	&-0.59079	&-0.0012	&0.2644	&14.77\\
J000730.59-004815.7	&1.87746615	&-0.80437922	&1	&3.8916	&11.1471	&16.69	&1.4257	&19.4255	&0.085269	&0.13012	&0.35178	&-1.8338	&-0.0197	&0.4258	&14.77\\
J034208.9-011448.2	&55.53708812	&-1.24674200	&1	&3.307	&5.0052	&12.81	&2.2921	&18.2684	&0.0011853	&0.043761	&-0.085353	&-0.58348	&-0.00076894	&0.1111	&14.79\\
J032305.27+002115.4	&50.77198928	&0.35429742	&1	&3.0148	&6.3899	&18.04	&2.7836	&18.4119	&0.006964	&0.91646	&-0.33345	&0.95463	&-0.0019	&0.2416	&14.79\\
J002901.48+002102.9	&7.25616975	&0.35082772	&1	&4.6257	&6.4413	&15.18	&2.2701	&18.6306	&-0.030459	&-0.040203	&0.0051457	&0.34502	&-0.0091	&0.076	&14.8\\
J005846.77-004506.5	&14.69489929	&-0.75182185	&0.97	&2.7363	&5.4847	&12.03	&1.9598	&18.1634	&-0.019812	&-0.053655	&-0.056628	&1.0168	&-0.0002218	&0.3852	&14.8\\
J013725.41+005838.4	&24.35587777	&0.97734793	&1	&3.9277	&5.5483	&12.99	&2.0655	&18.4488	&-0.034818	&0.051343	&-0.13581	&0.075314	&-0.0015	&0.1488	&14.8\\
J011612.78-000628.3	&19.05328708	&-0.10787758	&0.99	&4.2756	&5.288	&11.15	&1.8151	&18.442	&-0.0089351	&-0.0381	&-0.17492	&-0.47493	&-0.0009369	&0.0659	&14.81\\
J011708.94+010616.7	&19.28726213	&1.10464084	&0.93	&2.3468	&4.1957	&11.51	&2.4913	&17.7747	&-0.028059	&-0.020946	&-0.062099	&-0.46486	&0.0016	&0.1062	&14.81\\
J005800.46-001727.2	&14.50192231	&-0.29089488	&0.99	&2.8437	&5.1734	&12.25	&2.1472	&18.0879	&0.05988	&0.0097334	&0.22949	&-0.24307	&0.0014	&0.2776	&14.82\\
J213001.55-011351.4	&322.50648570	&-1.23094487	&1	&4.1145	&6.167	&14.76	&2.2492	&18.6188	&0.025221	&0.11569	&-0.055947	&-0.33062	&0.00049602	&0.0831	&14.82\\
J010401.73-004847		&16.00723838	&-0.81305974	&0.95	&3.318	&10.1002	&18.64	&1.8867	&19.2181	&-0.01182	&-0.11534	&-0.088332	&0.96319	&-0.007	&0.2942	&14.83\\
J233413.5+010148.4	&353.55626412	&1.03013378	&1	&3.2097	&6.1514	&14.77	&2.1454	&18.6026	&-0.03659	&-0.0080011	&-0.1193	&-0.15758	&-0.0021	&0.1956	&14.83\\
J223954.96-005919		&339.97904119	&-0.98861147	&1	&3.6077	&13.2744	&22.27	&1.905	&19.633	&0.011519	&0.35414	&-0.16742	&-1.0067	&-0.00051286	&0.2966	&14.84\\
J224544.74+010527.9	&341.43642301	&1.09108514	&1	&4.6055	&5.5938	&12.32	&1.8293	&18.4847	&-0.028146	&-0.061947	&0.15546	&0.89582	&0.0031	&0.1919	&14.85\\
J223045.48-010927.3	&337.68952754	&-1.15758815	&0.9	&3.6677	&12.8301	&18.77	&1.6141	&19.4409	&0.013797	&0.33252	&-0.16073	&-0.5191	&-0.0022	&0.43	&14.85\\
J024847.62-000633		&42.19842593	&-0.10917862	&0.86	&1.704	&4.307	&11.22	&2.6051	&17.4899	&0.071406	&0.0084674	&0.18868	&-0.15743	&0.00095301	&0.1972	&14.86\\
J001426.22+010350.8	&3.60928197	&1.06413569	&0.93	&1.9181	&4.2918	&11.17	&2.5473	&17.5994	&0.00067622	&-0.033878	&-0.032919	&-0.037764	&0.0071	&0.2072	&14.87\\
J015057.09+001404.2	&27.73789383	&0.23450808	&1	&3.226	&7.637	&16.57	&2.1531	&18.7384	&-0.03124	&0.057026	&-0.055255	&0.057246	&0.0089	&0.3795	&14.88\\
J002935.59-001406.6	&7.39829902	&-0.23517470	&1	&2.9263	&5.873	&15.92	&2.6701	&18.4369	&-0.026348	&-0.055164	&0.012083	&0.96878	&0.00042886	&0.145	&14.88\\
J222704.24+004517.5	&336.76769142	&0.75487411	&0.98	&3.302	&7.2755	&13.55	&1.7749	&18.6162	&-0.055876	&0.011286	&-0.27514	&-0.94251	&-0.0031	&0.3736	&14.88\\
J224223.94+011336.8	&340.59977329	&1.22691494	&1	&3.4292	&8.824	&15.78	&1.7966	&18.9445	&0.032081	&-0.18037	&-0.032733	&-0.83759	&0.0142	&0.3777	&14.89\\
J225510.03-002433.8 $^a$	&343.79180378	&-0.40939121	&1	&--- --- &--- ---	&--- ---	&--- ---	&--- ---	&--- ---		&--- ---		&--- ---		&--- ---		&--- ---	&--- ---	&--- ---\\
J024122.53+010640.8	&40.34388019	&1.11134030	&0.45	&1.403	&5.1375	&9.45	&1.8394	&17.6673	&-0.13034	&0.030038	&-0.32852	&-0.039536	&-0.0036	&0.4458	&14.92\\
J033845.73+011004.8	&54.69056321	&1.16802485	&1	&2.9388	&4.1105	&12.51	&2.8372	&17.8938	&-0.018453	&0.024214	&-0.066	&0.52901	&0.000054835	&0.0994	&14.94\\
J032528.7+001642.7	&51.36960070	&0.27853557	&0.97	&1.9696	&5.0688	&11.79	&2.4101	&17.7532	&-0.012671	&0.009981	&-0.044623	&0.13191	&0.0054	&0.38	&14.95\\
J031654.91-000231.1	&49.22881959	&-0.04199280	&1	&2.2112	&5.0854	&12.31	&2.3119	&17.8364	&0.059812	&-0.016033	&-0.030788	&-0.16606	&0.0055	&0.432	&14.96\\
  \noalign{\smallskip}\hline
\end{tabular}
\tablenotes{a}{0.95\textwidth}{The photometric results for J225510.03-002433.8 are not reliable, so it is not taken into the analyses afterwards.}
\end{table}

\label{lastpage}

\end{document}